\documentclass{article}
\usepackage{graphicx}
\usepackage{pdfpages,epstopdf}
\DeclareGraphicsExtensions{.pdf}
\newcommand{\be}{\begin{equation}}
\newcommand{\ee}{\end{equation}}
\newcommand{\bea}{\begin{eqnarray}}
\newcommand{\eea}{\end{eqnarray}}
\newcommand{\bfg}{\begin{figure}[p]}
\newcommand{\efg}{\end{figure}}
\newcommand{\ta}{\Theta}
\newcommand{\tp}{\Theta^{\prime}}
\newcommand{\tpp}{\Theta^{\prime \prime}}
\newcommand{\tppp}{\Theta^{\prime \prime \prime}}
\newcommand{\ph}{\ensuremath{\frac{\pi}{2}}}
\begin{document}
\bibliographystyle{plain}

  \title{The Geometry of Light Paths \\ for \\ 
         Equiangular Spirals}
  \author{E. Hitzer~\footnote{Dep. of Mech. Engineering, Fukui Univ., 
          Bunkyo 3-9-1, 910-8507 Fukui, Japan, hitzer@mech.fukui-u.ac.jp}}
  \date{1 December 1999 (rev. version)}
\maketitle

\begin{abstract}

First geometric calculus alongside its description of equiangular
spirals, reflections and rotations is introduced briefly. Then single
and double reflections at such a spiral are investigated. It proves
suitable to distinguish incidence from the \textit{right} and \textit{left} relative
to the radial direction. The properties of geometric light propagation
inside the equiangular spiral are discussed, as well as escape conditions and
characteristics. Finally the dependence of right and left incidence from the
source locations are examined, revealing a well defined inner \textit{critical}
curve, which delimits the area of purely right incident propagation. 
This critical curve is self similar to the original equiangular spiral.

\end{abstract}

\section{Introduction}

Deformations of circular discs lead to new promising laser resonators, 
dramatically improving output and beam quality~\cite{Phot-Bill,Laser-Flash}. In this paper
I want to look at the properties of \textit{spiral} deformations. 
In order to do this I will partly apply the very well suited geometric 
calculus as developed by D. Hestenes and G. Sobczyk~\cite{DH:NFII,DH:CAGC}.

\subsection{Prerequisites from Geometric Calculus}

Since I am interested in discs, I will only use a \textit{real} two 2-dimensional 
Euclidean vector space $\mathcal{E}_2$ to represent a plane and its \textit{real} 
plane geometric algebra $\mathcal{G}_2$. Fundamental for the notion vector in geometric
calculus is the associative geometric product of vectors {\boldmath $a$},{\boldmath $b$}:  
\be
 \mbox{\boldmath $a$} \mbox{\boldmath $b$} = 
 \mbox{\boldmath $a$} \cdot \mbox{\boldmath $b$} +
 \mbox{\boldmath $a$} \wedge \mbox{\boldmath $b$} 
\ee
composed of the conventional scalar inner product 
{$\mbox{\boldmath $a$} \cdot \mbox{\boldmath $b$}  $} and the outer product 
{$\mbox{\boldmath $a$} \wedge \mbox{\boldmath $b$} $}.
{$\mbox{\boldmath $a$} \wedge \mbox{\boldmath $b$} $} simply represents the oriented
area swept out by  {\boldmath $b$}, when displaced parallel along {\boldmath $a$} 
as seen in fig. \ref{fg:area-ab}. 
\bfg
%\vspace{2cm}
  %\begin{picture}(150,40)
    %\put(100,0){\scalebox{0.5}{\includegraphics[0in,1in][9in,9in]{ahatb}}}
		%\put(100,0){\scalebox{0.5}{\includegraphics{ahatb}}}
  %\end{picture}
 \begin{center}
   \scalebox{0.5}{\includegraphics{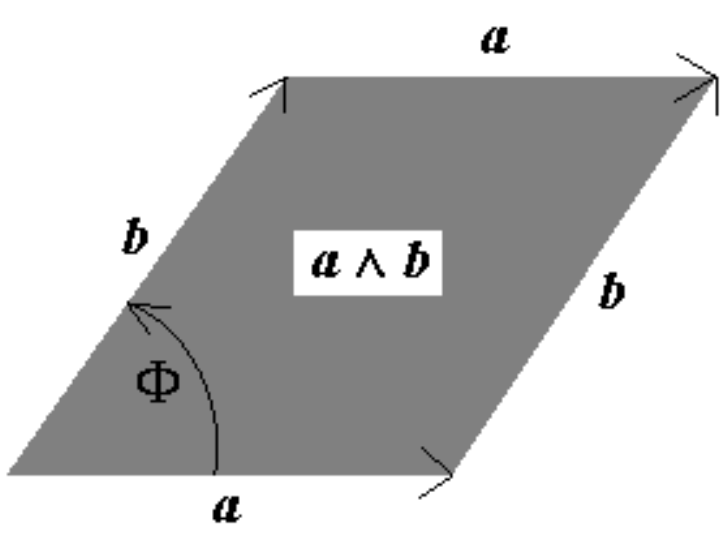}}
 \end{center}
 \caption{Oriented area {$\mbox{\boldmath $a$} \wedge \mbox{\boldmath $b$}$}. }
 \vspace{1cm}
 \label{fg:area-ab}
\efg
There are two orientations in following the outer contours of this area: \textit{plus} 
for going first along {\boldmath$a$} and then {\boldmath $b$}, etc. and \textit{minus} for
going first along {\boldmath$b$} and then {\boldmath $a$}, etc.

Hence, {$\mbox{\boldmath $a$} \wedge \mbox{\boldmath $b$} =
        - \mbox{\boldmath $b$} \wedge \mbox{\boldmath $a$} $}. 
{$\mbox{\boldmath $a$} \wedge \mbox{\boldmath $b$} $} is also 
called a bivector or in this case pseudoscalar, because its rank of 2 is maximal
in the plane geometric algebra $\mathcal{G}_2$ 
of scalars(1), vectors(2) and bivectors(1). 

The product of two vectors is a spinor.\footnote{Please compare \cite{DH:NFII}, p. 51 and 55
on the relation of spinors and complex numbers. It is important to note that the bivector \textbf{i} 
has algebraic and geometric properties beyond those of the traditional \textit{imaginary numbers}.} 
With the help of the oriented unit area element,
called ${\mathbf{i}} \in \mathcal{G}_2$  it can be written~\footnote{
Bold italic lowercase letters indicate vectors and nonbold italic lowercase letters 
indicate a vector`s length.} 
in exponential form:
\begin{eqnarray}
  \mbox{\boldmath $a$} \mbox{\boldmath $b$} & = &
   a \, b \, \, exp( {\mathbf{i}}  \Phi )          \nonumber \\
  a & = & \sqrt{\mbox{\boldmath $a$} \mbox{\boldmath $a$}}  
       =  \sqrt{\mbox{\boldmath $a$}^2}  
 \label{eq:gcexp}
 \\
  b & = & \sqrt{\mbox{\boldmath $b$}^2}  \nonumber 
\end{eqnarray}
with ${\mathbf{i}}^2 = -1$. This property of ${\mathbf{i}}$ can easily be
shown by chosing an orthonormal basis in $\mathcal{E}_2$: 
$\{ \mbox{\boldmath $\sigma_1$}, \mbox{\boldmath $\sigma_2$} \}$ with 
$\mbox{\boldmath $\sigma_1$}^2 = \mbox{\boldmath $\sigma_2$}^2=1;  
\mbox{\boldmath $\sigma_1$} \cdot \mbox{\boldmath $\sigma_2$} = 0$. ${\mathbf{i}}$ can than be written as 
${\mathbf{i}} 
   = \mbox{\boldmath $\sigma_1$} \mbox{\boldmath $\sigma_2$} 
   = \mbox{\boldmath $\sigma_1$} \wedge \mbox{\boldmath $\sigma_2$} $.
Hence  ${\mathbf{i}}^2 
  = (\mbox{\boldmath $\sigma_1$} \wedge \mbox{\boldmath $\sigma_2$} )
     \mbox{\boldmath $\sigma_1$} \mbox{\boldmath $\sigma_2$} 
  = -(\mbox{\boldmath $\sigma_2$} \wedge \mbox{\boldmath $\sigma_1$} )
      \mbox{\boldmath $\sigma_1$} \mbox{\boldmath $\sigma_2$}
  = - \mbox{\boldmath $\sigma_2$} \mbox{\boldmath $\sigma_1$} 
      \mbox{\boldmath $\sigma_1$} \mbox{\boldmath $\sigma_2$} 
  = - \mbox{\boldmath $\sigma_2$} \mbox{\boldmath $\sigma_2$} 
  = -1$.

Given that both $\mbox{\boldmath $a$}$ and $\mbox{\boldmath $b$}$ are unit vectors
\be
  \mbox{\boldmath $a$} \mbox{\boldmath $b$} = exp( {\mathbf{i}}  \Phi )          
\ee
can be used to describe the rotation of {\boldmath $a$} into {\boldmath $b$}:
\be
  \mbox{\boldmath $a$} \,\, exp( {\mathbf{i}}  \Phi ) 
  = \mbox{\boldmath $a$}\mbox{\boldmath $a$}\mbox{\boldmath $b$}
  = \mbox{\boldmath $a$}^2 \mbox{\boldmath $b$}
  = \mbox{\boldmath $b$}
\ee 

Further elements of geometric calculus will be introduced as needed throughout this
paper. 

\subsection{The Equiangular Spiral}

A circle of radius $x_0$ centered at the origin as shown in fig. 
\ref{fg:circle} may therefore be discribed as
\be
  \mbox{\boldmath $x$} = \mbox{\boldmath $x_o$} exp({\mathbf{i}}  \Phi); 
  \Phi \in \mathbf{R}.
\ee
\bfg
 %\vspace{2cm}
  %\begin{picture}(150,20)
    %\put(80,-10){\scalebox{0.5}{\includegraphics[0in,1in][9in,3in]{circle1}}}
		%\put(80,-10){\scalebox{0.5}{\includegraphics{circle1}}}
  %\end{picture} 
 \begin{center}
   \scalebox{0.5}{\includegraphics{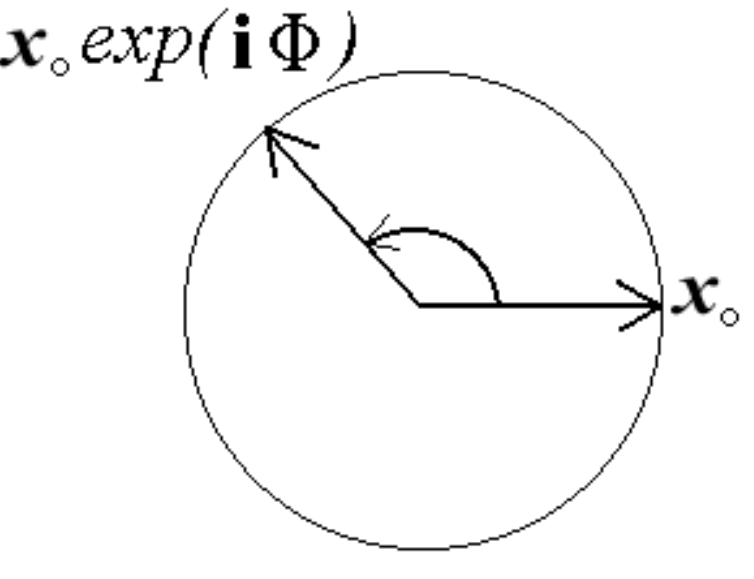}}
 \end{center}
 \caption{Circle of radius $x_0$.}
  \label{fg:circle}
\efg
In order to describe an equiangular spiral one needs to add a scalar multiple of 
$\Phi$ in the exponential:
\be
  \mbox{\boldmath $x$} = \mbox{\boldmath $x_o$} exp({\mathbf{i}}  \Phi + t \Phi  ); 
  \,\,\,\, t = const. \in \mathbf{R}.
  \label{eq:esdef}
\ee
The tangent to the curve $\mbox{\boldmath $x$}(\Phi)$ is defined as the 
first scalar derivative: 
\be
 \partial_{\Phi} \mbox{\boldmath $x$}  
                 = \mbox{\boldmath $x_o$} 
                   exp( ({\mathbf{i}}  + t) \Phi  ) ({\mathbf{i}}  + t) 
                 = \mbox{\boldmath $x$} {\mathbf{i}} + \mbox{\boldmath $x$} t
 \label{eq:estan}
\ee
The operation of ${\mathbf{i}}$ on $\mbox{\boldmath $x$}$ in (\ref{eq:estan}) is an 
anticlockwise rotation by a right angle as shown in fig. \ref{fg:estan}, since
\be
  exp({\mathbf{i}}  \frac{\pi}{2}) 
   = cos \frac{\pi}{2} + {\mathbf{i}} sin \frac{\pi}{2}
   = {\mathbf{i}}  
  \label{eq:iexp} 
\ee
%%%
%%%New separation command between floating figures!
%%%
\floatsep1cm
%%%
%%%
\bfg
 %\vspace{2cm}
  %\begin{picture}(150,80)
    %\put(80,0){\scalebox{0.4}{\includegraphics[0in,1in][9in,9in]{tangent1}}}
		%\put(80,0){\scalebox{0.4}{\includegraphics{tangent1}}}
  %\end{picture} 
	\begin{center}
	  \scalebox{0.4}{\includegraphics{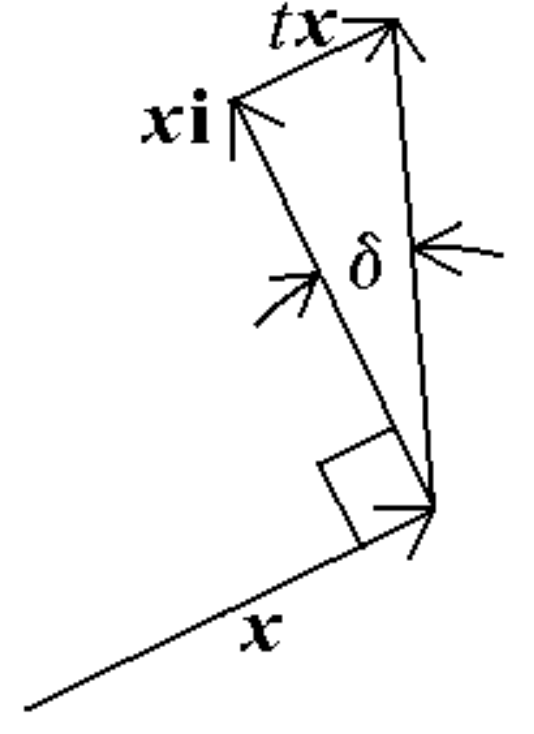}}
	\end{center}
  \caption{The tangent vector of the equiangular spiral.}
  \label{fg:estan}
\efg
It immediately follows that
\be
  \mid \mbox{\boldmath $x$} {\mathbf{i}} \mid = x \mbox{  and  } 
  \mid t \mbox{\boldmath $x$}  \mid = t x
  \label{eq:estangleC}
\ee
and therefore
\be
  tan \delta = \frac{\mid t \mbox{\boldmath $x$}  \mid }
                   {\mid \mbox{\boldmath $x$} {\mathbf{i}} \mid }
             = t \mbox{  or  } \delta = tan^{-1}{}t\,\,(=arctan{}\,\, t)
\ee
Hence the tangent $\partial_{\Phi} \mbox{\boldmath $x$}$ has relative to {\boldmath $x$}
the angle 
\be
  \frac{\pi}{2} + \delta = \frac{\pi}{2} + tan^{-1}{}t
  \label{eq:estangle}
\ee
The deviation of this angle from the case of a pure circle is $\delta = tan^{-1}{}t$
independent of $\Phi$. That is the reason, why the spiral (\ref{eq:esdef}) is called 
\textit{equiangular}~\footnote{Comp. \cite{DH:NFII}, p. 155.}.

Let me assume for the rest of this paper, without loss of generality, that 
$t>0.$~\footnote{\label{fn:tszero} $t<0$ would just mean that I would have to interchange the later 
defined notions of incidence from the right and from the left. This is trivial. In eq. 
(\ref{eq:estangleC}) I have already quietly made this assumption for t.}
Such an equiangular spiral with $\delta = 0.1 (\approx t)$ is shown in fig.
\ref{fg:espiral}.
\bfg
  %\begin{picture}(150,80)
    %\put(90,0){\scalebox{0.4}{\includegraphics[0in,1in][9in,9in]{spiral1}}}
		%\put(90,0){\scalebox{0.4}{\includegraphics{spiral1}}}
  %\end{picture} 
	\begin{center}
	  \scalebox{0.4}{\includegraphics{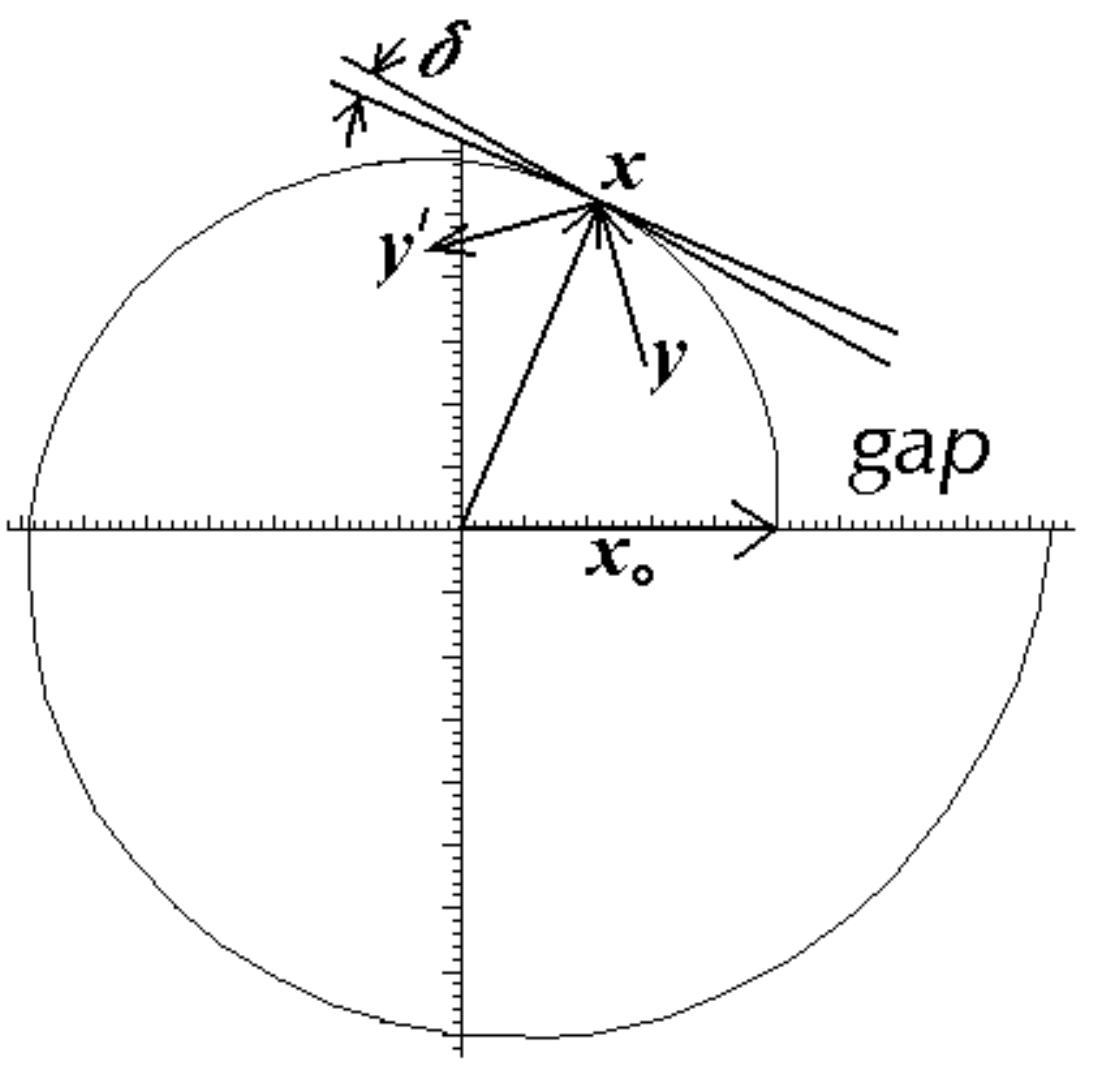}}
	\end{center}
  \caption{The equiangular spiral for $\delta = 0.1$. }
  \label{fg:espiral}
\efg

\section{Reflections at an Equiangular Spiral}

\subsection{Single Reflections \label{ssc:sgref}}

In order to describe a reflection at the equiangular spiral we need to know
the unit normal {\boldmath $n$} at any point $\mbox{\boldmath $x$}(\Phi)$. For
a circle the unit normal points in the same direction as the radius vector: 
$\mbox{\boldmath $n$}=\mbox{\textit{{\textbf {\^{x}}}}}$. 
For an equiangular spiral the tangent vector has the angle $\frac{\pi}{2} + \delta$
relative to {\textit{\textbf {\^{x}}}}. {\boldmath $n$} will hence be equal to 
{\textit{\textbf {\^{x}}}} rotated clockwise by $\delta$:
\be
  \mbox{\boldmath $n$} = \mbox{\textit{\textbf {\^{x}}}} exp(-\mathbf{i} \delta)
  \label{eq:esnorm}
\ee
Every vector $\mbox{\boldmath $y$} $ incident at a point $\mbox{\boldmath $x$} $
of the equiangular spiral (e.g. representing a light ray to be reflected at 
$\mbox{\boldmath $x$} $) can be uniquely decomposed into components parallel and
perpendicular relative to {\boldmath $n$}:
\begin{eqnarray}
  \mbox{\boldmath $y$} & = & \mbox{\boldmath $y$}_{\|} + \mbox{\boldmath $y$}_{\perp} 
  \mbox{  with  } \nonumber \\
  \mbox{\boldmath $y$}_{\|} & = & \mbox{\boldmath $y$} \cdot \mbox{\boldmath $n$}  
                                  \mbox{\boldmath $n$} 
  \label{eq:vdec} \\
  \mbox{\boldmath $y$}_{\perp} & = & \mbox{\boldmath $y$} - \mbox{\boldmath $y$}_{\|}
                                 = \mbox{\boldmath $y$} \mbox{\boldmath $n$}  
                                  \mbox{\boldmath $n$} 
                                 - \mbox{\boldmath $y$} \cdot \mbox{\boldmath $n$}  
                                  \mbox{\boldmath $n$} 
                                 = (\mbox{\boldmath $y$} \mbox{\boldmath $n$}   
                                 - \mbox{\boldmath $y$} \cdot \mbox{\boldmath $n$} ) 
                                  \mbox{\boldmath $n$} 
                                 = \mbox{\boldmath $y$} \wedge \mbox{\boldmath $n$} 
                                   \mbox{\boldmath $n$}  
  \nonumber
\end{eqnarray}
Here I used the convention that indicated inner and outer products should be performed
before an adjacent geometric product (comp.~\cite{DH:CAGC}, p. 7).

The reflection will then be described by
\begin{eqnarray}
  \mbox{\boldmath $y$}^{\prime} & = & -\mbox{\boldmath $n$} 
                                      \mbox{\boldmath $y$} \mbox{\boldmath $n$}
\stackrel{(\ref{eq:vdec})}{=} -\mbox{\boldmath $n$}
  (\mbox{\boldmath $y$} \cdot \mbox{\boldmath $n$} \mbox{\boldmath $n$} 
    + \mbox{\boldmath $y$} \wedge \mbox{\boldmath $n$} \mbox{\boldmath $n$}  ) 
  \mbox{\boldmath $n$}  
\nonumber \\
& = & 
  (- \mbox{\boldmath $y$} \cdot \mbox{\boldmath $n$} \mbox{\boldmath $n$} 
    - \mbox{\boldmath $n$}  \mbox{\boldmath $y$} \wedge \mbox{\boldmath $n$} ) 
   \mbox{\boldmath $n$}^2 
  \stackrel{\mbox{\boldmath $n$}^2=1 }{=} 
  - \mbox{\boldmath $y$}_{\|} + \mbox{\boldmath $y$}_{\perp} 
  \label{eq:esref}
\end{eqnarray} 
In the last step one uses equ. (\ref{eq:vdec}) again and the fact that
$-\mbox{\boldmath $n$}\mbox{\boldmath $y$} \wedge \mbox{\boldmath $n$}
= -\mbox{\boldmath $n$} \wedge (\mbox{\boldmath $y$} \wedge \mbox{\boldmath $n$})
  -\mbox{\boldmath $n$} \cdot (\mbox{\boldmath $y$} \wedge \mbox{\boldmath $n$})
= (\mbox{\boldmath $y$} \wedge \mbox{\boldmath $n$}) \cdot \mbox{\boldmath $n$} 
= (\mbox{\boldmath $y$} \wedge \mbox{\boldmath $n$}) \mbox{\boldmath $n$} $.
This holds true, because in the geometric algebra $\mathcal{G}_2$ the outer
product is associative as well and the outer product of two equal vectors is always zero:
$-\mbox{\boldmath $n$} \wedge (\mbox{\boldmath $y$} \wedge \mbox{\boldmath $n$})
= -\mbox{\boldmath $n$} \wedge \mbox{\boldmath $y$} \wedge \mbox{\boldmath $n$}
= \mbox{\boldmath $n$} \wedge \mbox{\boldmath $n$} \wedge \mbox{\boldmath $y$}
= 0$~\footnote{In this case one could argue alternatively, that in the \textit{plane}
geometric algebra $\mathcal{G}_2$ no 3-dim. volumes exist and the outer product
of 3 vectors therefore always vanishes.}.
The anticommutativity of the vector {\boldmath $n$} and the 
bivector $\mbox{\boldmath $y$} \wedge \mbox{\boldmath $n$}$ follows from the 
general definition of the inner product between vectors {\boldmath $v$} 
and bivectors $B$ (comp.~\cite{DH:CAGC}, p. 7):
\be
  \mbox{\boldmath $v$} \cdot B = \frac{1}{2} (\mbox{\boldmath $v$} B 
                                              - B \mbox{\boldmath $v$} )
\ee
With the help of (\ref{eq:esnorm}) equation (\ref{eq:esref}) can be rewritten as
\be
  \mbox{\boldmath $y$}^{\prime} = 
    -\mbox{\textit{\textbf {\^{x}}}} exp(-\mathbf{i} \delta)
        \mbox{\boldmath $y$} 
     \mbox{\textit{\textbf {\^{x}}}} exp(-\mathbf{i} \delta)
  = exp(\mathbf{i} \delta) 
(-\mbox{\textit{\textbf {\^{x}}}} \mbox{\boldmath $y$} 
     \mbox{\textit{\textbf {\^{x}}}}) exp(-\mathbf{i} \delta)
  \label{eq:esref2} 
\ee
The last equality follows from the fact that in $\mathcal{G}_2$ as we just saw, 
vectors and bivectors anticommute, i.e.
\be
  \mbox{\textit{\textbf {\^{x}}}} \mathbf{i} 
  = - \mathbf{i} \mbox{\textit{\textbf {\^{x}}}}
\ee
and from expanding $exp(-\mathbf{i} \delta)$ in powers of $-\mathbf{i} \delta$.
The inner bracket 
$(-\mbox{\textit{\textbf {\^{x}}}} \mbox{\boldmath $y$} \mbox{\textit{\textbf {\^{x}}}}) 
= \mbox{\boldmath $y$}^{\prime}_c $
represents according to eq. (\ref{eq:esref}) a reflection at a circle with radius 
vector {\textit{\textbf {{x}}}}. The two spinors attached to the left and 
to the right can now both be moved to one side, e.g. as
\be
  \mbox{\boldmath $y$}^{\prime} = \mbox{\boldmath $y$}^{\prime}_c 
  exp(- 2 \mathbf{i} \delta)
\ee
$\mbox{\boldmath $y$}^{\prime} $ is now understood as the composition of the reflection
at the circle with radius vector {\textit{\textbf {{x}}}} and a clockwise rotation
by the angle of $2 \delta$. 
According to (\ref{eq:esref2}) $\mbox{\boldmath $y$}^{\prime}$ may also be written as
\be
  \mbox{\boldmath $y$}^{\prime} = R^{\dag}{}(-\delta) \mbox{\boldmath $y$}^{\prime}_c 
                                  R(-\delta)
  \mbox{   with   } R(-\delta) = exp(-\mathbf{i} \delta)
  \label{eq:esRR}
\ee
Here the ${}^{\dag}$ operator is the reversion operator, i.e. reversing every geometric
product in $R(-\delta)$. That $R^{\dag}{}(-\delta) = R(\delta)$ can easily be seen from 
the fact that 
$\mathbf{i}^{\dag} 
= (\mbox{\boldmath $\sigma_1$}\mbox{\boldmath $\sigma_2$})^{\dag} 
= \mbox{\boldmath $\sigma_2$}\mbox{\boldmath $\sigma_1$}
= - \mbox{\boldmath $\sigma_1$}\mbox{\boldmath $\sigma_2$}
= - \mathbf{i}$.
Spinors like $R(-\delta)$ are also called rotors, because they elegantly describe
rotations. (\ref{eq:esRR}) is the double sided spinorial description of rotations 
(comp.~\cite{DH:NFII}, p. 277 ff.).

The increase in angle by $2 \delta$ is a clear consequence of the fact, that the 
tangent of the equiangular spiral is tilted by the constant angle $\delta$ relative
to the tangent of a circle with radius vector {\textit{\textbf {{x}}}}.

\subsection{Two Successive Reflections \label{us:2ref}}

The interesting question to ask is what is the angular difference 
$\Delta \Phi = \Phi^{\prime} - \Phi$ 
between two successive reflections. This will also result in an answer to the 
nontrivial question of how the incident angle for a second successive reflection 
$\Theta^{\prime \prime}$ depends on the incident angle of the first $\Theta$ as 
shown in fig.~\ref{fg:2ref}. 
(For a circle we would just have 
$\Theta = \Theta^{\prime} = \Theta^{\prime \prime} = \Theta^{\prime \prime \prime}$
and $\Delta \Phi = \pi - 2 \Theta$.)
The treatment of this problem naturally splits in the two cases of the second 
reflection occuring 
before $(\Phi^{\prime} > \Phi)$ or after $(\Phi^{\prime} < \Phi)$ 
crossing\footnote{With a ray \textit{crossing the gap} I mean that this ray intersects 
with the line segment between the origin and $\mbox{\boldmath $x$}_{2\pi}$. The point of 
intersection may either be between the origin and $\mbox{\boldmath $x$}_0$ or between 
$\mbox{\boldmath $x$}_0$ and $\mbox{\boldmath $x$}_{2\pi}$. In the first case, the
ray will continue to be reflected inside the equiangular spiral, wheras in the second case it will
escape through the gap. This distinction is made in detail in section~\ref{uus:escape}.}
 the gap at $\Phi = 2 \pi$, 
i.e. the line segment between $\mbox{\boldmath $x$}_0$ and $\mbox{\boldmath $x$}_{2\pi}$.
In the following I will distinguish \textit{incidence from the right} 
for which the reflected ray leaves the reflecting boundary of the equiangular spiral to the left of the radius 
vector of the point of reflection, i.e. 
$\tp = \ta + 2 \delta \in [0,\ph + \delta]$,
and
\textit{incidence from the left} for which the reflected ray leaves to the right of the radius vector, i.e.
$\tp \in [-\ph + \delta,0[$.
First, incidence from the right will be treated in detail. It entails the possibility of rays escaping 
through the gap as will soon be shown. Furthermore incidence from the left can be viewed as the 
reverse situation of incidence from the right, excluding the possibility of escape. 
\bfg
%  \vspace{2cm}
  %\begin{picture}(150,100)
    %\put(80,0){\scalebox{0.5}{graphics[0in,1in][9in,9in]{tworefl1}}}
		%\put(80,0){\scalebox{0.5}{\includegraphics{tworefl1}}}
  %\end{picture} 
	\begin{center}
	  \scalebox{0.5}{\includegraphics{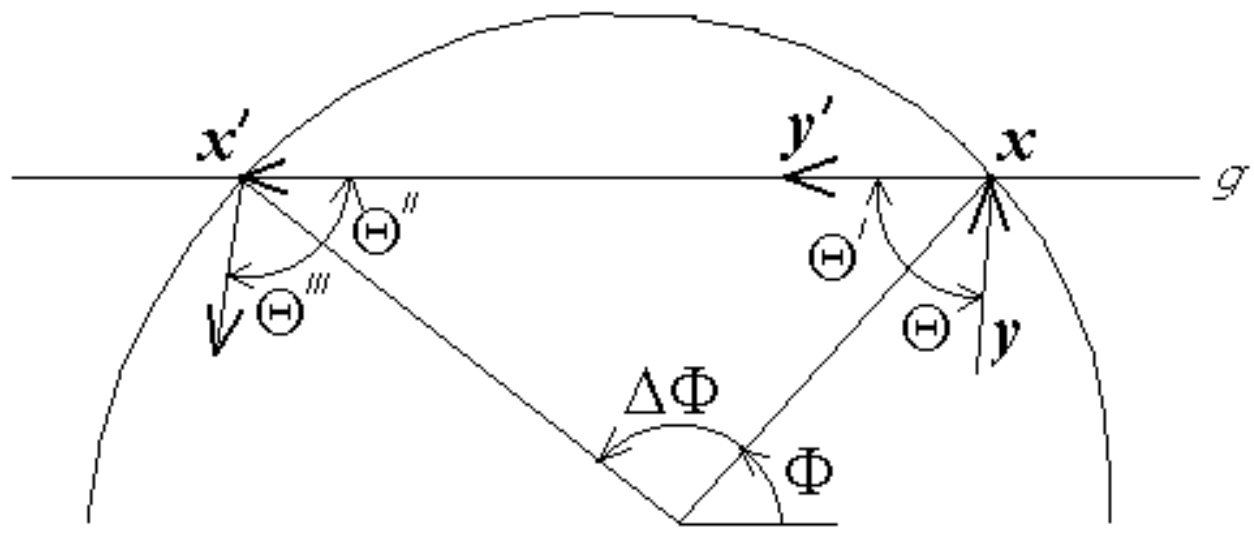}}
	\end{center}
  \caption{Two successive reflections.}
  \label{fg:2ref}
\efg

\subsubsection{Reflections Without Crossing the Gap}

A unit vector of incidence at angle $\Theta$ relative to 
$\mbox{\textit{\textbf {\^{x}}}}$ is given by 
\be
  \mbox{\textit{\textbf {\^{y}}}} 
  = \mbox{\textit{\textbf {\^{x}}}} exp(\mathbf{i} \Theta)
\ee 
The reflected vector is according to (\ref{eq:esRR}) 
\bea
  \mbox{\textit{\textbf {\^{y}}}}^{\prime} & = & 
   R^{\dag}{}(-\delta) 
   (-\mbox{\textit{\textbf {\^{x}}}} 
         \mbox{\boldmath $y$} \mbox{\textit{\textbf {\^{x}}}})
   R(-\delta)
  =
  exp(\mathbf{i} \delta ) 
  (-\mbox{\textit{\textbf {\^{x}}}} \mbox{\textit{\textbf {\^{x}}}} 
  exp(\mathbf{i} \Theta) \mbox{\textit{\textbf {\^{x}}}})
  exp(-\mathbf{i} \delta ) 
  \nonumber \\
  & = &
  - exp(\mathbf{i} (\delta + \Theta ))\mbox{\textit{\textbf {\^{x}}}}  
  exp(-\mathbf{i} \delta )
  = \mbox{\textit{\textbf {\^{x}}}} exp(-\mathbf{i} (2 \delta + \Theta + \pi ))
\eea
In the last step I used the operator identity~\footnote{Compare also the explanations
after eq. (\ref{eq:gcexp}) and before eq. (\ref{eq:iexp}).}
$-1 = exp(\pm \mathbf{i} \pi )$,
i.e. a rotation by $\pi$.

In order to find the location of the second reflection 
$\mbox{\textit{\textbf x}}^{\prime}(\Phi^{\prime})$ 
I construct a straight line $g$ through
{\boldmath $x$} with direction $\mbox{\textit{\textbf {\^{y}}}}^{\prime} $.
$\mbox{\textit{\textbf x}}^{\prime}(\Phi^{\prime})$ 
will be its (i.e. the other) point of intersection with the equiangular spiral.
\bea
  g : \mbox{\textit{\textbf y}}^{\prime} & = &
      \mbox{\textit{\textbf x}} + \lambda \mbox{\textit{\textbf {\^{y}}}}^{\prime}
  =
  \mbox{\boldmath $x_o$}  exp(\mathbf{i} \Phi) exp (t \Phi)
   + \lambda \mbox{\textit{\textbf {\^{x}}}}_{\mathbf{0}} 
     exp(\mathbf{i}(\Phi - \Theta - 2 \delta -\pi)) 
  \nonumber \\
 & \stackrel{!}{=}&
  \mbox{\boldmath $x_o$} exp(\mathbf{i} \Phi^{\prime}) exp (t \Phi^{\prime})
 \label{eq:rcon1}
\eea
This may be rewritten as
\be
  \mbox{\boldmath $x_o$} exp(\mathbf{i}(\Theta + 2 \delta )) 
  - \lambda \mbox{\textit{\textbf {\^{x}}}}_{\mathbf{0}} 
 = \mbox{\boldmath $x_o$} exp(t(\Phi^{\prime}- \Phi) 
   exp(\mathbf{i}(\Phi^{\prime} - \Phi + \Theta + 2 \delta ))
\ee
Multiplication with 
$\mbox{\boldmath $x_o$}^{-1} = \frac{\mbox{\boldmath $x_o$}}{\textstyle x_0^2}$
from the left gives:
\be
exp(\mathbf{i} \Theta^{\prime}) - \frac{\lambda}{x_0} 
= exp(t \Delta \Phi) exp(\mathbf{i}(\Delta \Phi + \Theta^{\prime}))
  \label{eq:rcon1a}
\ee
where $\Theta^{\prime} = \Theta + 2 \delta$ as we have seen in section 
\ref{ssc:sgref}.

Equation (\ref{eq:rcon1a}) has scalar and bivector parts, which must be satisfied
separately. The bivector part divided by $\mathbf{i}$ reads
\be
sin\Theta^{\prime} = exp(t \Delta \Phi) sin(\Delta \Phi + \Theta^{\prime})
  \label{eq:rcon1b}
\ee
(\ref{eq:rcon1b}) is a transcendental equation for 
$\Phi^{\prime} = \Delta \Phi + \Phi$ which may either be solved graphically 
or numerically with Newton iteration. For the graphical solution it is convenient
to multiply (\ref{eq:rcon1b}) with $exp(t \Theta^{\prime})$:
\be
  f(\Delta \Phi + \Theta^{\prime}) 
  := exp(t(\Delta \Phi + \Theta^{\prime})) sin(\Delta \Phi + \Theta^{\prime})
  = exp(t \Theta^{\prime}) sin(\Theta^{\prime})
  =: f(\Theta^{\prime})
  \label{eq:rcon1c}            
\ee
$f$ is a simple sinus function with an exponential envelope 
$exp(t \Theta^{\prime})$ depending on the parameter $t = tan \delta$.
(For the circle t = 0 and therefore $exp(t \Theta^{\prime}) \equiv 1$.)
\bfg
  %\begin{picture}(150,150)
    %\put(20,0){\scalebox{0.6}{\includegraphics[0in,1in][9in,9in]{gsolI1}}}
		%\put(20,0){\scalebox{0.6}{\includegraphics{gsolI1}}}
  %\end{picture}
	\begin{center}
	  \scalebox{0.6}{\includegraphics{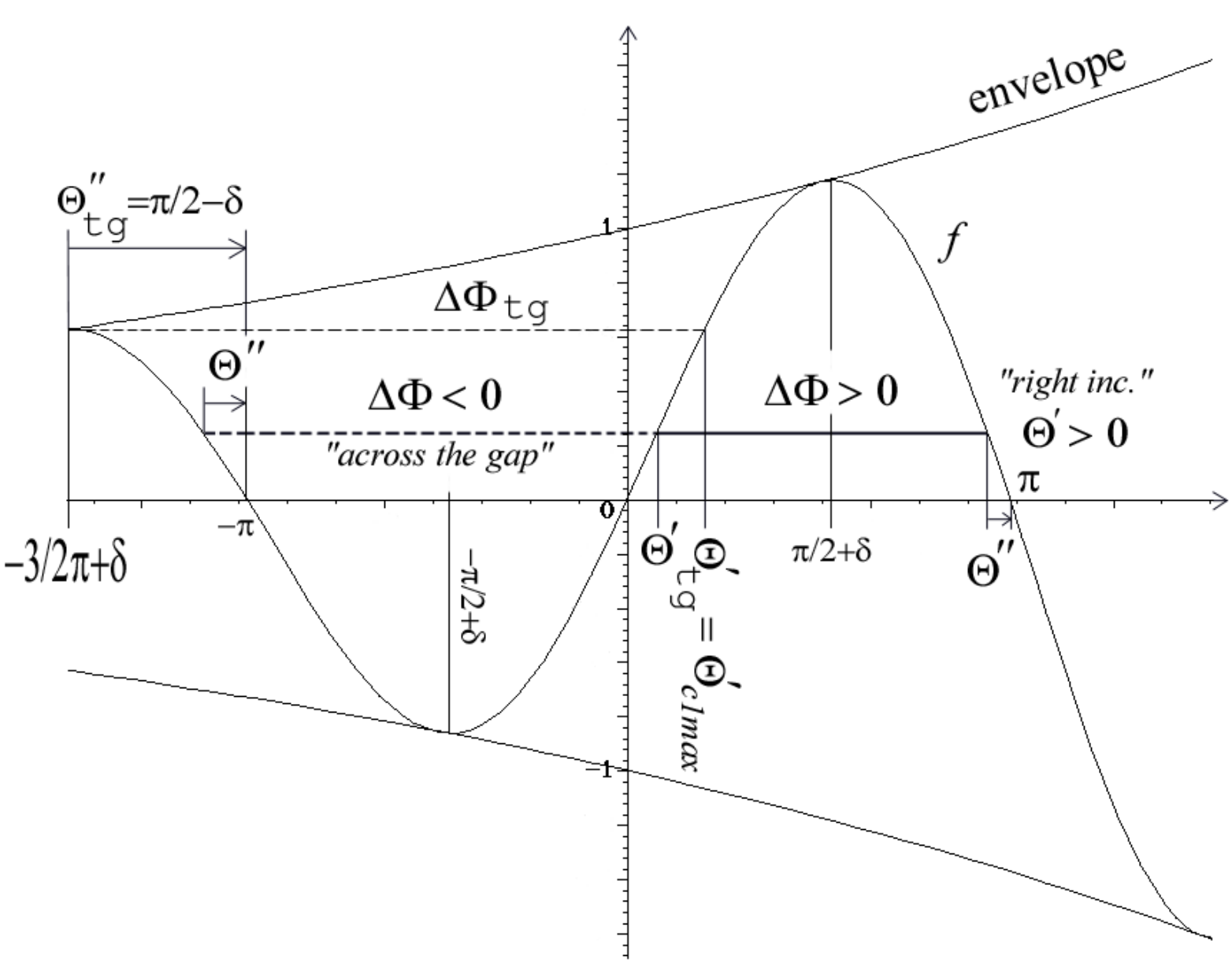}}
	\end{center}
 %\vspace{2cm}
 \caption{Graphical solution for successive reflections of right incident ray.}
 \label{fg:gsol1}
\efg  

For the circle we have $\Theta^{\prime} = \Theta^{\prime \prime}$.
Yet for the equiangular spiral the upper right half of 
fig.~\ref{fg:gsol1} clearly shows that 
$\Theta^{\prime} \geq \Theta^{\prime \prime}$, since the envelope 
$exp(t \Theta^{\prime})$ is a monotone increasing function of $\Theta^{\prime}$.
It is however possible to show, that 
\be
  \Theta  < \Theta^{\prime \prime} 
  \mbox{   for   } \Theta \in [-2 \delta, \frac{\pi}{2}- \delta[.
  \label{eq:ttpcon}
\ee

Before proving this let me first establish another useful property of the $\tpp (\ta)$ 
dependence:
\be
  \frac{\partial \Theta^{\prime \prime}}{\partial \Theta} > 0
  \mbox{   for   }  \Theta \in [-2 \delta, \frac{\pi}{2}- \delta[.
  \label{eq:ttpder}  
\ee
which shows that $\tpp$ is a monotone increasing function of $\ta$
in $\Theta \in [-2 \delta, \frac{\pi}{2}- \delta[$. 

\subsubsection*{Proof of Eq. (\ref{eq:ttpder}) }

Since according to fig.~\ref{fg:2ref} $\tpp = \pi - (\Delta \Phi + \tp)$ we may
write eq. (\ref{eq:rcon1c}) as
\be
   \mbox{\textit{\~{h}}}(\tpp) := exp(t(\pi-\tpp)) sin \tpp 
   = exp(t(\ta + 2 \delta)) sin(\ta + 2 \delta) =: \mbox{\textit{\~{f}}}(\ta) 
  \label{eq:rcon1d}
\ee
Therefore 
\be
 \frac{\partial \tpp}{\partial \ta} 
 = \frac{\partial \mbox{\textit{\~{f}}}}{\partial \ta}
   \left(\frac{\partial \mbox{\textit{\~{h}}}}{\partial \tpp} \right)^{-1}
  \label{eq:ipder1}
\ee

Differentiation of {\textit{\~{f}}} and {\textit{\~{h}}} respecively gives
\bea
\frac{\partial \mbox{\textit{\~{f}}}}{\partial \ta} 
  & = &
  \mbox{\textit{\~{f}}}(\ta) (tan \delta + \frac{1}{tan(\ta + 2 \delta)} )
  \nonumber \\
\frac{\partial \mbox{\textit{\~{h}}}}{\partial \tpp} 
  & = &
  \mbox{\textit{\~{h}}}(\tpp) (-tan \delta + \frac{1}{tan \tpp} )
\eea
According to (\ref{eq:ipder1}) and (\ref{eq:rcon1d}) this results in
\be
  \frac{\partial \Theta^{\prime \prime}}{\partial \Theta} 
  = \frac{1 + \frac{1}{tan(\ta + 2 \delta) tan \delta} }{
    -1 + \frac{1}{tan \tpp {}tan \delta } }
  \label{eq:ipder2}
\ee
The numerator of the rhs. of eq. (\ref{eq:ipder2}) equals
\be
  \frac{1 + tan^2{}\delta}{tan(\ta + \delta) + tan^2{}\delta} 
> 0 \mbox{   for   } \Theta \in [-2 \delta, \frac{\pi}{2}- \delta[,
\ee
whilst the supremum value~\footnote{Equation (\ref{eq:rcon1d}) 
and its diagram fig.~\ref{fg:gsol1} show that 
$\tpp \in [0, \frac{\pi}{2}- \delta[$ 
for $\Theta \in [-2 \delta, \frac{\pi}{2}- \delta[$.} 
of $tan \tpp$ in the nominator of the same equation can
be written as
\be
  tan \tpp_{\mathit{sup}} = tan(\frac{\pi}{2} - \delta) = \frac{1}{tan \delta}.
  \label{eq:invtan}
\ee
Hence $tan \tpp tan \delta < 1$ and
$-1 + \frac{1}{tan \tpp {}tan \delta } > 0$
for $\Theta \in [-2 \delta, \frac{\pi}{2}- \delta[$
and $\tpp \in [0, \frac{\pi}{2}- \delta[$ respectively. 
This conludes the proof of eq. (\ref{eq:ttpder}) since now both the numerator
and the nominator of eq. (\ref{eq:ipder2}) are proven to be positive 
for $\Theta \in [-2 \delta, \frac{\pi}{2}- \delta[$
and $\tpp \in [0, \frac{\pi}{2}- \delta[$ respectively. 

Let me finally remark that for the case of tangential incidence 
(at the inside of the equiangular spiral) with 
$\ta = \tpp = \frac{\pi}{2}-\delta$ we obviously have $\Delta \Phi = 0$.

\subsubsection*{Proof of Eq. (\ref{eq:ttpcon}): $\Theta  < \Theta^{\prime \prime}$}

For the other extreme of $\ta = -2 \delta$
and $\tp = \ta + 2 \delta = 0$ we have according to eq. (\ref{eq:rcon1d}) and
figs.~\ref{fg:2ref} and \ref{fg:gsol1} respectively, $\tpp = 0$, i.e.
\be
  \ta < \tpp \mbox{   for   }  \ta = -2 \delta
  \label{eq:0cond}
\ee
Based on this, the most important
step of the proof will be to first show that $\ta \neq \tpp$ for
$\Theta \in [-2 \delta, \frac{\pi}{2}- \delta[$. Because $\tpp$ is a continuous 
differentiable function of $\Theta$ in this interval, it will then immediately follow
from eq. (\ref{eq:0cond}) that $\Theta  < \Theta^{\prime \prime}$ for all
$\Theta \in [-2 \delta, \frac{\pi}{2}- \delta[$.
That $\tpp (\ta)$ is continuous and differentiable is evident from the 
analytic expression for $\frac{\partial \Theta^{\prime \prime}}{\partial \Theta} $ 
given in eq. (\ref{eq:ipder2}), which is well defined for 
$\Theta \in [-2 \delta, \frac{\pi}{2}- \delta[$. 

It now remains to show, that $\ta \neq \tpp$ for 
$\Theta \in [-2 \delta, \frac{\pi}{2}- \delta[$. 
I will start out supposing the opposite, i.e. $\ta = \tpp$ for some
$\Theta \in [-2 \delta, \frac{\pi}{2}- \delta[$ and show that this 
leads to contradictions. Supposing that for some 
$\Theta \in [-2 \delta, \frac{\pi}{2}- \delta[$ I would have $\ta = \tpp$, 
I can rewrite eq. (\ref{eq:rcon1b}) as
\be
  exp(t(\pi-\ta)) sin\ta - exp(t(\ta + 2 \delta)) sin(\ta + 2 \delta) = 0
  \label{eq:tetcon}
\ee
I will examine the validity of (\ref{eq:tetcon}) first in the intervall 
$\Theta \in [-2 \delta, 0[$, then at the point $\ta = 0$ and finally for
$\Theta \in ]0, \frac{\pi}{2}- \delta[$. 

First for $\Theta \in [-2 \delta, 0[$ we have $exp(t(\pi-\ta)) sin\ta < 0$ and
$exp(t(\ta + 2 \delta)) sin(\ta + 2 \delta) \geq 0$. Therefore the expression on the 
lhs. of eq. (\ref{eq:tetcon}) will be
\be
  exp(t(\pi-\ta)) sin\ta - exp(t(\ta + 2 \delta)) sin(\ta + 2 \delta) < 0
  \mbox{   for   } \Theta \in [-2 \delta, 0[
\ee
which is in contradiction to eq. (\ref{eq:tetcon}). 
Hence we have 
\be
  \ta \neq \tpp \mbox{   for   } \Theta \in [-2 \delta, 0[
  \label{eq:tneqt1}
\ee

Second for $\ta = 0$ the lhs. of eq. (\ref{eq:tetcon}) becomes
\be
  - exp(t 2 \delta) sin 2 \delta \neq 0 \mbox{   for   } 0 < \delta < \frac{\pi}{2}
\ee
in contradiction to eq. (\ref{eq:tetcon}). Hence I conlude again, that 
\be
  \ta \neq \tpp \mbox{   for   } \Theta = 0
  \label{eq:tneqt2}
\ee
(I am not concerned~\footnote{See also footnote~\ref{fn:tszero} on page 
\pageref{fn:tszero}.} 
with $\delta = 0$, i.e. the circle or $\delta = \frac{\pi}{2}$,
i.e. a straight half line 
in the direction of $\mbox{\textit{\textbf {\^{x}}}}_{\mathbf{0}}$  
beginning at {\boldmath $x_o$}, nor with $\delta > \frac{\pi}{2}$, which is
equivalent to an equiangular spiral winding clockwise with
$\delta^{\prime} = \pi - \delta$.)

Finally for $\Theta \in ]0, \frac{\pi}{2}- \delta[$ I will calculate the first
derivative of the lhs. of eq. (\ref{eq:tetcon}) with respect to $\ta$:
\bea
&&\hspace{-1cm}
\frac{\partial}{\partial \ta}
\left(exp(t(\pi-\ta)) sin\ta - exp(t(\ta + 2 \delta)) sin(\ta + 2 \delta) \right)
\nonumber \\
& = & - exp(t(\pi-\ta)) (t\, sin \ta - cos \ta) 
\nonumber \\
&&    - exp(t(\ta + 2 \delta)) ( t\, sin(\ta + 2 \delta) + cos (\ta + 2 \delta) )
\nonumber \\
& \stackrel{t=\frac{sin \delta}{cos \delta}}{=} &
      - \frac{exp(t(\pi-\ta))}{cos \delta} ( sin \delta \,\, sin \ta - cos \delta \,\, cos \ta) 
\nonumber \\
&&
      - \frac{exp(t(\ta + 2 \delta)) }{cos \delta} 
        ( sin \delta \,\, sin(\ta + 2 \delta) + cos \delta \,\, cos (\ta + 2 \delta) )
\nonumber \\
& = & \frac{1}{cos \delta}
       (exp(t(\pi-\ta))cos(\ta + \delta) - exp(t(\ta + 2 \delta))cos(\ta + \delta) )
\nonumber \\
& = & \frac{cos(\ta + \delta)}{cos \delta} exp(t(\ta + 2 \delta))
       (exp(t(\pi - 2 (\ta + \delta))) - 1 ) 
\nonumber \\
& > & 0 \,\,\,\, \mbox{   for   } \Theta \in ]0, \frac{\pi}{2}- \delta[
\eea
The last inequality holds, because both 
$\frac{cos(\ta + \delta)}{cos \delta} exp(t(\ta + 2 \delta)) > 0$ and
$(\pi - 2 (\ta + \delta)) > 0 $ hold for $\Theta \in ]0, \frac{\pi}{2}- \delta[$.
I therefore conlude, that the lhs. of eq. (\ref{eq:tetcon}) is a stricly monotone
increasing function of $\ta$. This means that it will not vanish for any 
$\Theta \in ]0, \frac{\pi}{2}- \delta[$ because at the supremum of 
$\ta_{\mathit{sup}} = \frac{\pi}{2}-\delta$ of the intervall 
$]0, \frac{\pi}{2}- \delta[$ the lhs. of eq. (\ref{eq:tetcon}) (and its first
derivative with respect to $\ta$) is actually zero. The nonvanishing of the lhs. of
eq. (\ref{eq:tetcon}) for $\Theta \in ]0, \frac{\pi}{2}- \delta[$ contradicts
eq. (\ref{eq:tetcon}) and hence the supposition $\ta = \tpp$ for 
any $\Theta \in ]0, \frac{\pi}{2}- \delta[$. I conlude that 
\be
  \ta \neq \tpp \mbox{   for   } 
  \Theta \in ]0, \frac{\pi}{2}- \delta[
  \label{eq:tneqt3}
\ee

Taking eqs. (\ref{eq:tneqt1}), (\ref{eq:tneqt2}) and (\ref{eq:tneqt3}) together it is 
shown that $\ta \neq \tpp$ for $\Theta \in [-2 \delta, \frac{\pi}{2}- \delta[$. 
This conludes the prove for eq. (\ref{eq:ttpcon}) as argued above.

%%%
%%%NEW INSERTED TEXT:
%%%
Indeed, eq. (\ref{eq:ttpcon}) holds not only for $\tp \in [0, \frac{\pi}{2}+ \delta[$
and reflections which don`t cross the gap radius between the origin and $\mbox{\boldmath $x$}_{2\pi}$, 
but also if the gap radius is crossed by a ray. The left side of fig.~\ref{fg:gsol1} will be
explained in section~\ref{uus:acgap}. Anticipating this and refering in addition to fig.~\ref{fg:across1}, 
fig.~\ref{fg:gsol1} shows that because of the strictly monotone character of the exponential envelope
$\tpp>\tp (=\ta+2\delta)$ holds. Hence we immediately end up with
\be
   \tpp > \ta
   \nonumber
\ee 
for right incident rays travelling across the gap as well. Therefore eq.~(\ref{eq:ttpcon}) 
holds in full generality for any right incident ray.

The property $\Theta  < \Theta^{\prime \prime} $ physically means that successive
reflections of light rays incident from the right bend the paths of these rays
closer and closer to the tangential direction of 
$\partial_{\Phi} \mbox{\boldmath $x$}$, i.e. to the equiangular spiral disc boundary.

After infering all these properties from eq. (\ref{eq:rcon1b}) a comment on its
familiar geometric meaning is in place. Because 
\bea
  sin(\Delta \Phi + \tp) & = & sin(\pi - (\Delta \Phi + \tp) ) = sin \tpp 
  \mbox{   and   }
  \nonumber \\
  exp(t\Delta \Phi) & = & \frac{exp(t\Phi^{\prime}) }{exp(t\Phi) } 
                      = \frac{x^{\prime}}{x}
  \nonumber
\eea
we can rewrite eq. (\ref{eq:rcon1b}) as
\be
  x^{\prime} sin \tpp = x sin \tp 
  \label{eq:rcon1tr} 
\ee
Looking at the angles in fig.~\ref{fg:2ref} we see that eq. (\ref{eq:rcon1b}) 
in the form of eq. (\ref{eq:rcon1tr}) 
embodies nothing else but the familiar law of sinuses of the triangle formed
by the side vectors {\boldmath $x$} and $\mbox{\boldmath $x$}^{\prime}$.

\subsubsection{Reflecting Across the Gap \label{uus:acgap}}

What happens now, if a refelcted ray actually crosses the $\Phi = 2 \pi$ radius, 
i.e. the line segment between the origin and $\mbox{\boldmath $x$}_{2\pi}$ 
part of which forms the gap?
\bfg
  %\begin{picture}(150,200)
    %\put(15,0){\scalebox{0.5}{\includegraphics[0in,1in][9in,9in]{gapref1}}}
		%\put(15,0){\scalebox{0.5}{\includegraphics{gapref1}}}
  %\end{picture}
	\begin{center}
	  \scalebox{0.5}{\includegraphics{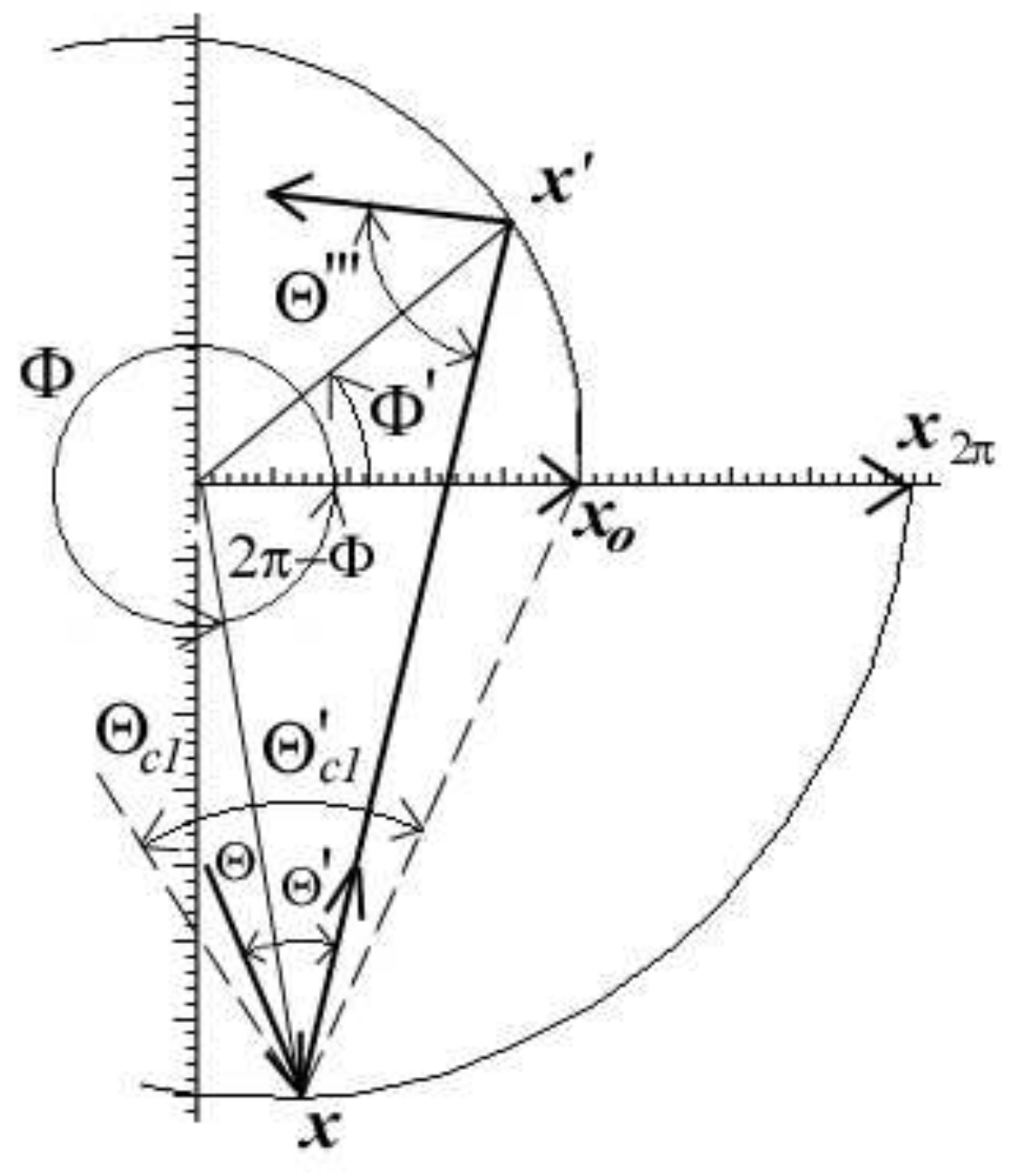}}
	\end{center}
  %\vspace{2cm}
  \caption{Successive reflections across the gap.}
  \label{fg:across1}
\efg
As may readily be seen from fig.~\ref{fg:across1}, a ray passing the angle 
$\Phi = 2 \pi$ may continue to be reflected inside the equiangular spiral, 
for angles of incidence from the right of less than a critical angle $\ta_{c1}$,
which remains to be determined. Or it may eventually emanate from the spiral for
$\ta > \ta_{c1}$.

Eq.~(\ref{eq:rcon1b}) holds for this case equally. The only difference is that we now have to 
deal with a negative $\Delta\Phi=\Phi^{\prime}-\Phi$, since $\Phi^{\prime} < \Phi$. 
In fig.~\ref{fg:gsol1} this case is represented by the dashed line extending from $f(\tp)$ to the left. 
Where this line intersects $f$ again, somewhere in the intervall 
$\Delta\Phi + \tp \in [-\pi,-\frac{3}{2}\pi + \delta[$ we have $f(\Delta\Phi + \tp )$. The negative length of 
this horizontal line segment is precisely $\Delta\Phi$. The sum of angles in the triangle formed by the 
origin, $\mbox{\boldmath $x$}(\Phi)$ and $\mbox{\boldmath $x$}(\Phi^{\prime})$ gives
\be
    \tpp+\tp+\Delta\Phi = - \pi.
\ee
Hence the representation of $\tpp$, the angle of incidence at $\Phi^{\prime}$, as drawn in 
fig.~\ref{fg:gsol1}. 

\subsubsection{The Condition of Escape \label{uus:escape}}

\bfg
  %\begin{picture}(150,130)
    %\put(60,0){\scalebox{0.4}{\includegraphics[0in,1in][9in,9in]{escapecon1}}}
		%\put(60,0){\scalebox{0.4}{\includegraphics{escapecon1}}}
  %\end{picture} 
	\begin{center}
	  \scalebox{0.4}{\includegraphics{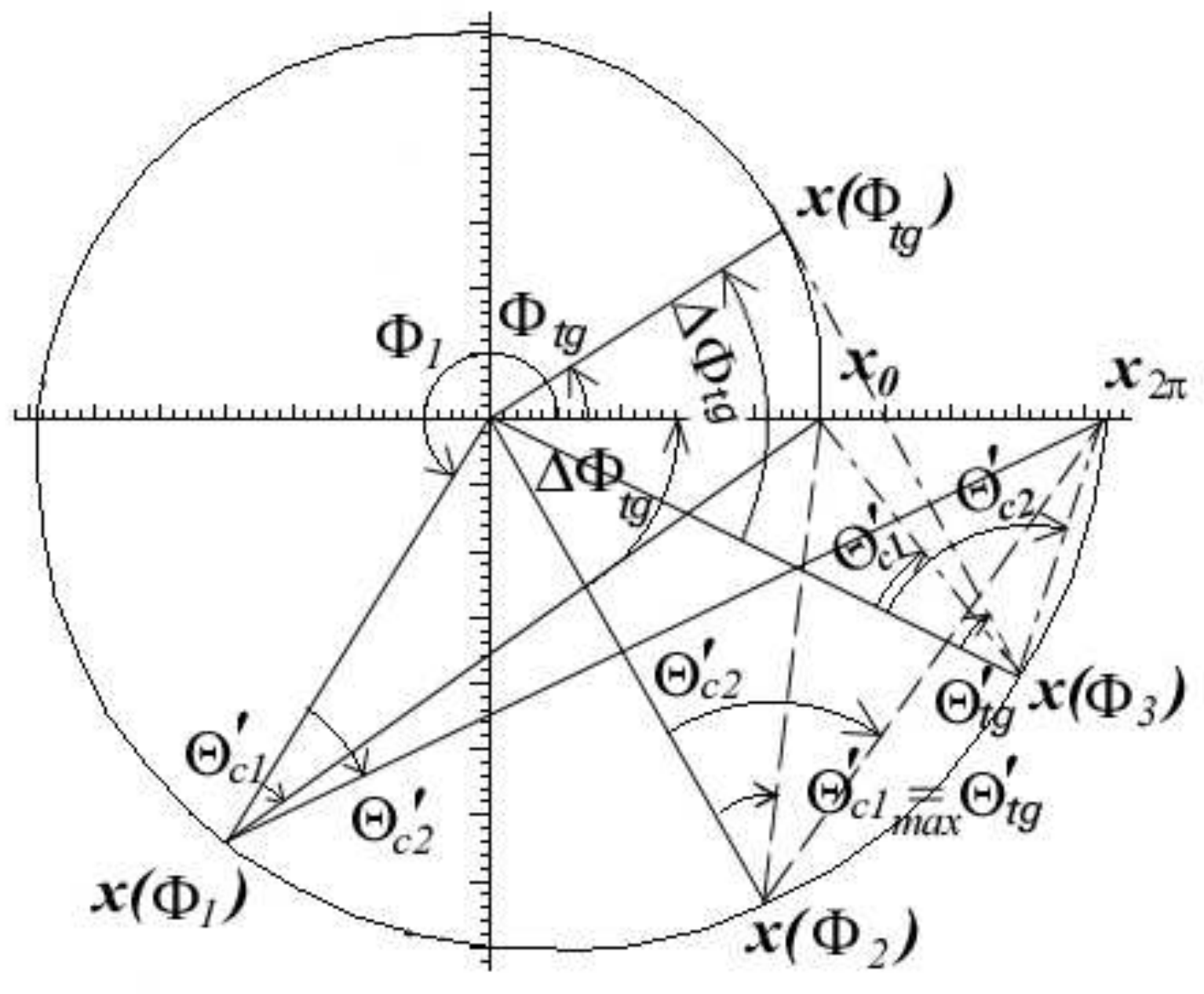}}
	\end{center}
 %\vspace{2cm}
  \protect\caption{Angular conditions\protect~\protect\footnotemark of escape.}
  \label{fg:esc1}
\efg

First for $0<\tp < \tp_{c1}(\Phi)$ a light ray will continue to go round anticlockwise
inside the spiral. 

Second for 
\be
  \tp_{c1}(\Phi) < \tp < \tp_{c2}(\Phi) 
  \label{eq:stesc}
\ee
 a light ray will leave 
the spiral straight away (comp. Fig.~\ref{fg:esc1}).

And third for $\tp > \tp_{c2}(\Phi)$ a ray will experience further reflections
between $\mbox{\boldmath $x$}(\Phi)$ and $\mbox{\boldmath $x$}(2 \pi)$ along
the equiangular spiral boundary until it finally
leaves the equiangular spiral without reentering the anticlockwise polygonal motion
of the rays of the first case. 
%
%The following passage needs to be further researched!
%
%The reason is eq. (\ref{eq:ttpder}) 
%$\frac{\partial \Theta^{\prime \prime}}{\partial \Theta} > 0$ together with the 
%observation from fig.~\ref{fg:esc1} that $\tp_{c2}$ (as well as $\tp_{c2}$) is
%a monotone decreasing function of $\Phi$ the location of the first reflection 
%
%

$\tp_{c1}(\Phi)$ (second incidence at {\boldmath $x_0$}) can be determined from 
eq. (\ref{eq:rcon1b}) by inserting $\Phi^{\prime} = 0$:
\be
  exp(t( - \Phi + \tp_{c1})) sin(- \Phi + \tp_{c1}) = exp(t\tp_{c1} ) sin \tp_{c1} .
  \label{eq:c1con}
\ee
Likewise $\tp_{c2}(\Phi)$ (second incidence at $\mbox{\boldmath $x$}(2 \pi)$)
can be determined from eq. (\ref{eq:rcon1b}) by inserting $\Phi^{\prime} = 2 \pi$:
\be
  exp(t(2 \pi - \Phi +\tp_{c2})) sin(2 \pi - \Phi +\tp_{c2}) 
  = exp(t\tp_{c2} ) sin \tp_{c2} .
  \label{eq:c2con}
\ee
Eq. (\ref{eq:c1con}) can be used to graphically determine $\tp_{c1}$ with the 
help~\footnotetext{
The $\Delta \Phi_{tg}$ corresponds to the $\Delta \Phi_{tg}$ as given in 
fig.~\ref{fg:gsol1} and described on page~\pageref{outer-inc}.}  
of fig.~\ref{fg:gsol1}: Set $\Delta \Phi = - \Phi$ for a given $\Phi$. 
For $\Delta \Phi < -\pi $ we fit a (dashed) horizontal line segment of 
length $\Delta \Phi $ 
between $f(\tp)$ to the right and $f(\Delta \Phi +\tp)$ to the left. The 
$\tp$, where the segment touches the curve $f(\tp)$ is $\tp_{c1}$. (The corresponding
cricital angle of incidence at $\mbox{\boldmath $x$}(\Phi)$ is 
$\ta_{c1} = \tp_{c1} - 2 \delta$.)  

By similarly fitting an (unbroken) horizontal line segment 
of length $\Delta \Phi = 2 \pi - \Phi > 0$ 
between $f(\tp)$ to the left and $f(\Delta \Phi +\tp)$ to the right in fig.~\ref{fg:gsol1} we can determine
$\tp_{c2}$ (and $\ta_{c2}$ respectively).

For rays which fulfil the condition of straight escape (\ref{eq:stesc}), one can 
further distinguish, whether they reflect a last time at the outer side of the 
equiangular spiral or not. This is essential for deciding the asymptotic directions
of light rays leaving the gap of the equiangular spiral.

The criterion for this may also be derived from eq. (\ref{eq:rcon1b}). Tangential 
incidence at the outer side of the equiangular spiral occurs for $\tp_{c1} < \tp_{tg}$
with $\tpp_{tg} = \frac{\pi}{2} - \delta$. The point of tangential incidence on the
outer side at $\Phi_{tg}$ can be determind graphically from fig.~\ref{fg:gsol1}. 
$\Delta \Phi_{tg} = \Phi_{tg} - \Phi <0 $ \label{outer-inc}is given by the 
horizontal distance between the maximum of $f$
to the left (at $\tpp_{tg} = \frac{\pi}{2} - \delta$) and $f(\tp)$ to the right. The
point on $f(\tp)$, where this (dashed) horizontal line ends is $\tp_{tg}$. 
In case that $\tp_{c1} < \tp_{tg}$ the intervall $\tp_{c1} < \tp < \tp_{c2}$ can
be subdivided into $\tp_{c1} < \tp < \tp_{tg}$, 
for which rays leaving straight will reflect a last time at the outer side of 
the spiral and $\tp_{tg} < \tp < \tp_{c2}$, for which rays leave straight without
further incidence at the outside.

The latter kind of rays will actually leave the gap under an angle 
$\Phi_e$ relative to the direction
of $\mbox{\textit{\textbf {\^{x}}}}_{\mathbf{0}}$ given by 
\be
  \Phi_e = \Phi - \pi - \tp
\ee 
and an abscissa of escape $x_e$ given by
\be
  x_e = x \frac{sin \tp}{sin \Phi_e}
\ee
as can be seen from fig.~\ref{fg:esdat}.
\bfg
  %\begin{picture}(150,150)
    %\put(70,0){\scalebox{0.4}{\includegraphics[0in,1in][9in,9in]{escapechar1}}}
		%\put(70,0){\scalebox{0.4}{\includegraphics{escapechar1}}}
  %\end{picture} 
	\begin{center}
	  \scalebox{0.4}{\includegraphics{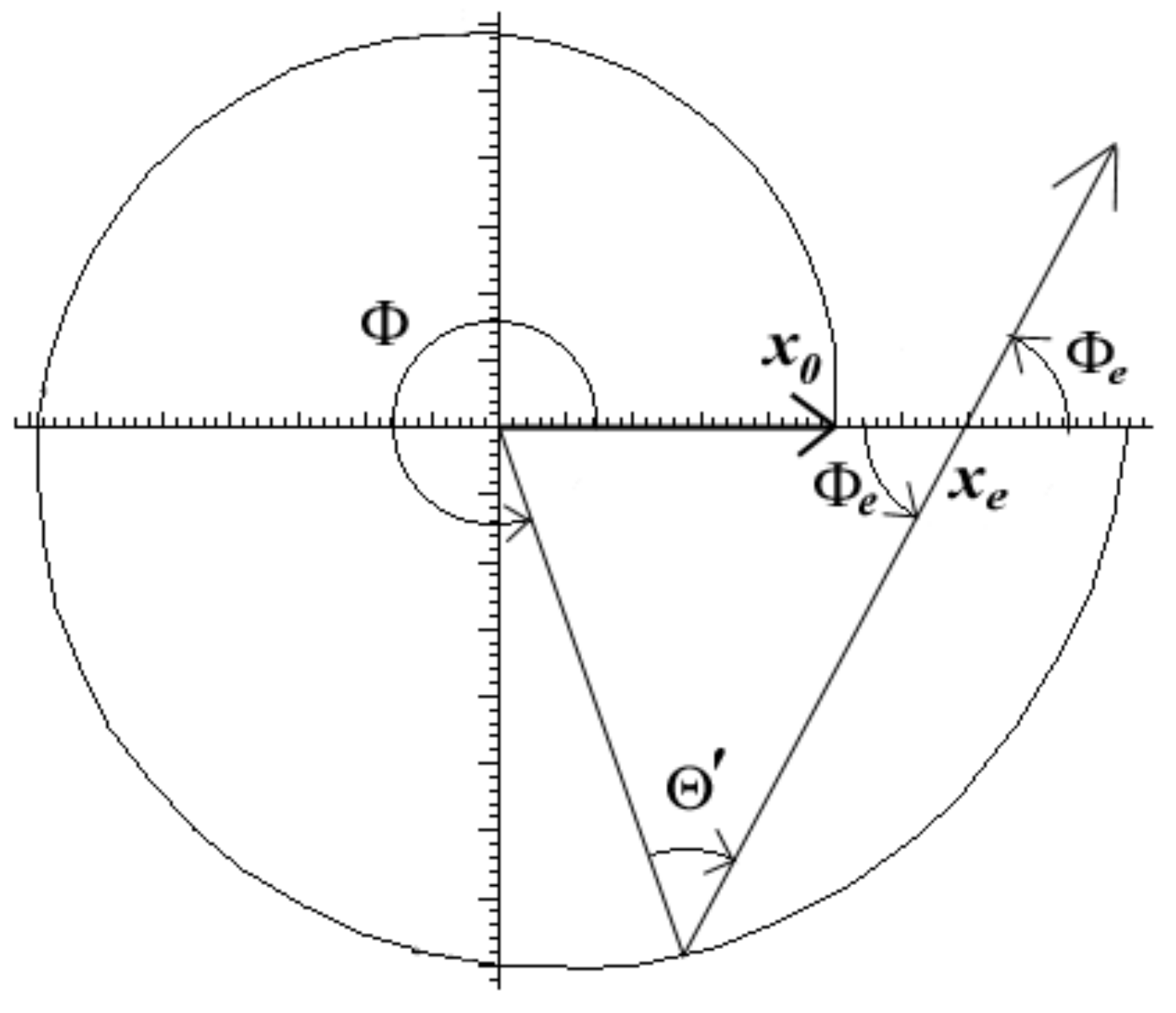}}
	\end{center}
% \vspace{2cm}
  \caption{Angle and abscissa of escape.}
  \label{fg:esdat}
\efg
This actually characterizes the asymptotic radiation completely in case that one
ignores the reflection of rays at the outside, e.g. because of absorption.

\subsection{Incidence from the Left}

As for incidence from the left, i.e. $0 > \tp \geq -\frac{\pi}{2} + \delta$,  
eq.~(\ref{eq:rcon1b}) now describes the graph of 
$f$ in the lower part of fig.~{\ref{fg:gsol1} 
with negative values of the ordinate. This is redrawn in fig.~\ref{fg:gsol2}. 

The angular position of $\mbox{\textit{\textbf{x}}}_{0}$ relative to $\mbox{\textit{\textbf{x}}}$
as illustrated in fig.~\ref{fg:leftin} tells whether a ray will cross the radial position $\Phi=2\pi$ of
the gap or not. For $\tp < \tp_{cl}$ the ray reflected at 
$\mbox{\textit{\textbf{x}}}(\Phi)$ will not cross the gap and
the unbroken horizontal line segment extending between $f(\tp)$ 
to the right and $f(\tp + \Delta\Phi)$ to the left applies. For $\tp > \tp_{cl}$ the reflected ray
will cross the gap and the dashed 
horizontal line segment extending between $f(\tp)$ to the left and $f(\tp+\Delta\Phi)$ to the right applies.
\bfg
  %\begin{picture}(150,110)
    %\put(90,0){\scalebox{0.3}{\includegraphics[0in,1in][9in,9in]{leftincid1}}}
		%\put(90,0){\scalebox{0.3}{\includegraphics{leftincid1}}}
  %\end{picture} 
	\begin{center}
	  \scalebox{0.3}{\includegraphics{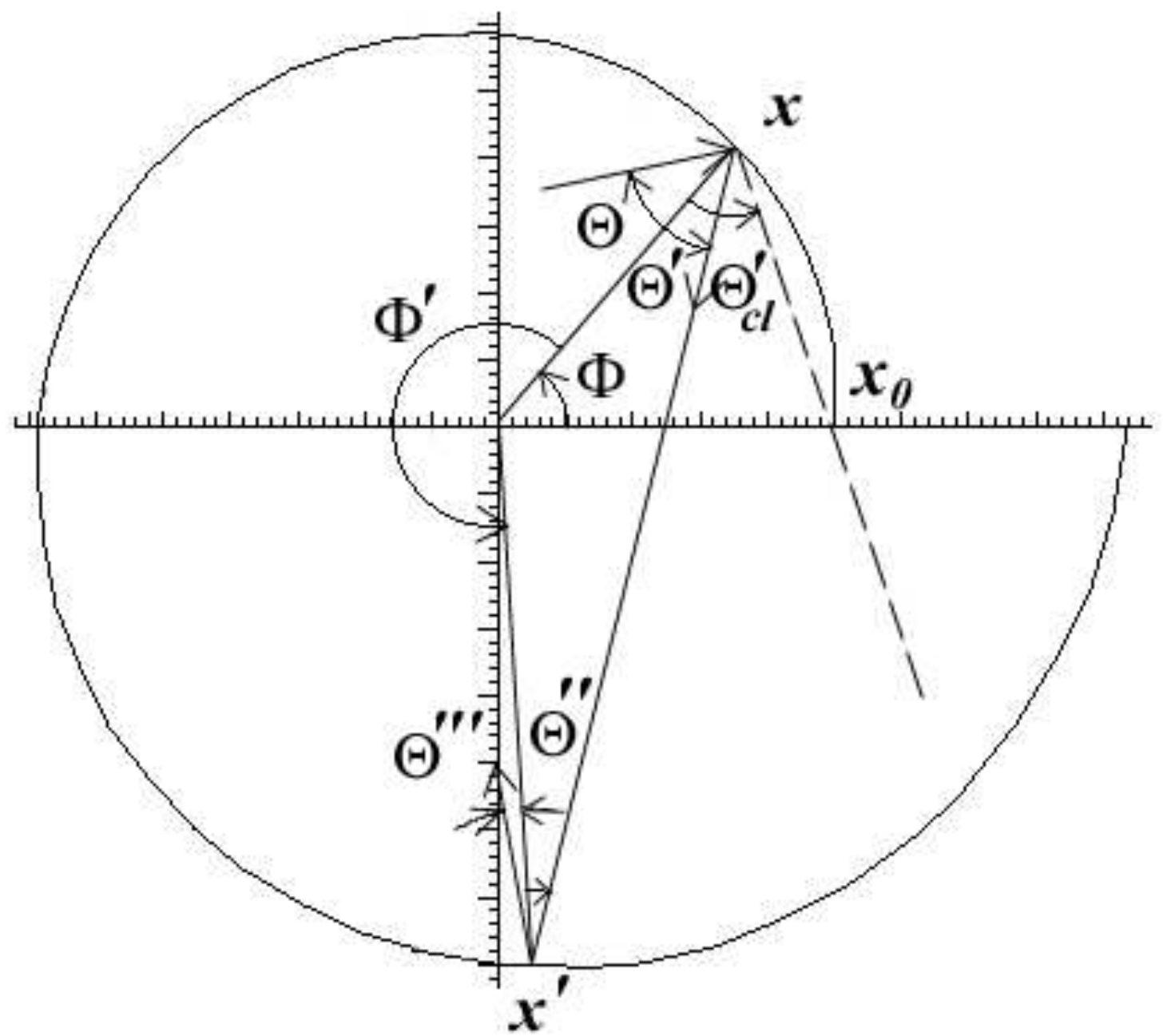}}
	\end{center}
% \vspace{2cm}
  \caption{Incidence from the left.}
  \label{fg:leftin}
\efg 
\bfg
  %\begin{picture}(150,140)
    %\put(30,0){\scalebox{0.4}{\includegraphics[0in,1in][9in,9in]{gsolII1}}}
		%\put(30,0){\scalebox{0.4}{\includegraphics{gsolII1}}}
  %\end{picture} 
	\begin{center}
	  \scalebox{0.4}{\includegraphics{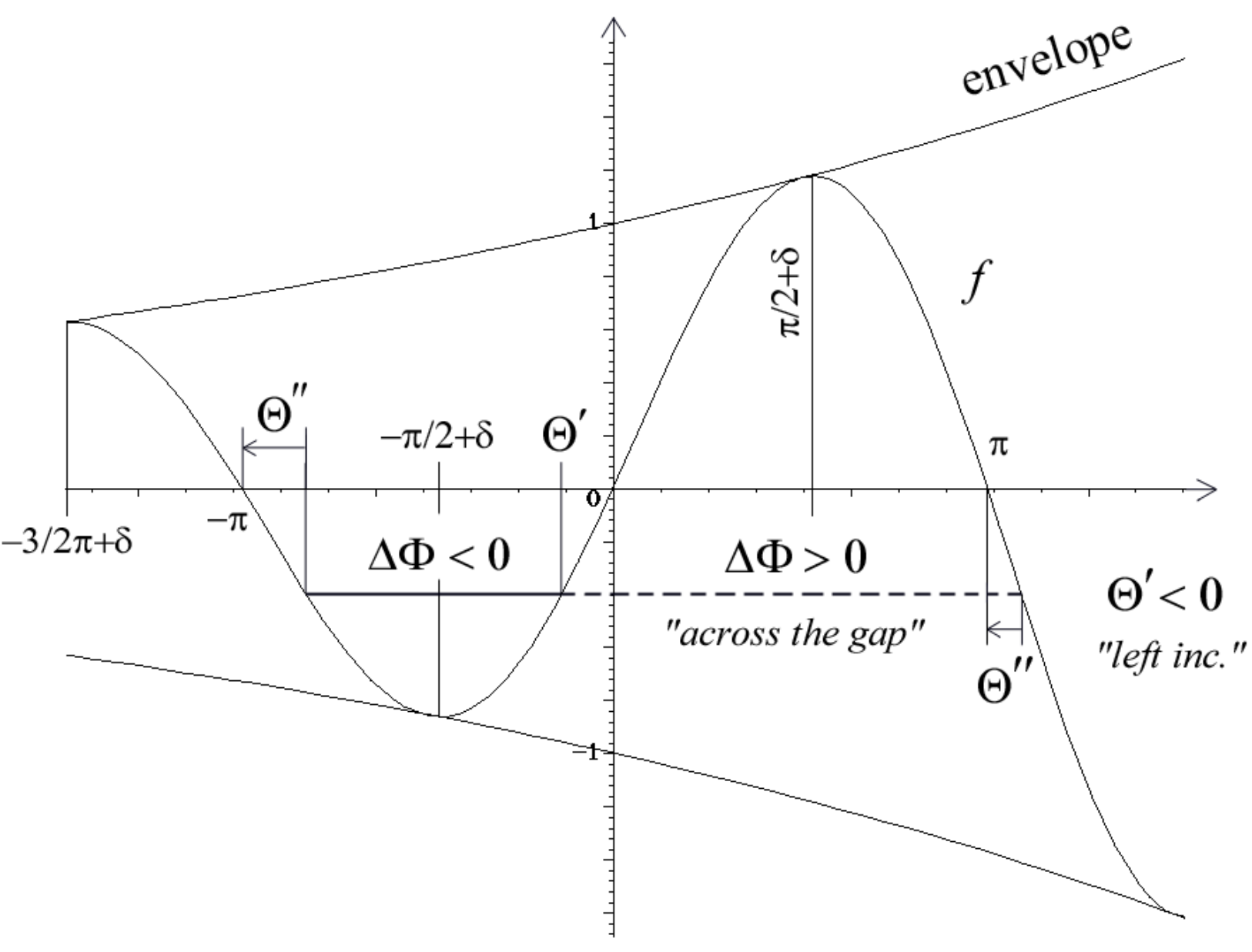}}
	\end{center}
% \vspace{2cm}
  \caption{Graphical solution for successive reflections of left incident ray.}
  \label{fg:gsol2}
\efg 

Another way to view rays which are incident from the left is as the reverse of corresponding rays 
incident from the right. According to fig.~\ref{fg:2ref} this would mean to mutually exchange $\Phi$ and
$\Phi^{\prime}$, $\ta$ and $\tppp$, and $\tp$ and $\tpp$:
\be
   \Phi \leftrightarrow  \Phi^{\prime}; 
   \ta \leftrightarrow \tppp;
   \tp \leftrightarrow \tpp.
   \label{eq:exlr}
\ee
Applying this to the inequality (\ref{eq:ttpcon}) as well we get
\be
  \tpp - 2 \delta = \tppp  <  \tp = \ta - 2 \delta 
    \mbox{   for   } \tp \in [0, \ph - \delta [ 
  \nonumber 
\ee
or adding $2 \delta$
\be
  \tpp  <  \ta \mbox{   for   } \ta \in [2 \delta, \ph + \delta [ 
\ee
This means that light rays with incidence from the left will be bent closer and closer 
to the radial direction \textit{\textbf{\^{x}}} as they perform a clockwise polygonal 
motion through the equiangular sprial. 
Once $\tpp$ has become smaller than $2 \delta$ they can
be treated as rays with incidence from the right with 
$\Theta \in [-2 \delta, \ph - \delta [$ 
as described earlier in section ~\ref{us:2ref}. In particular, as long as a ray remains in the
`incidence from the left' state, it will not leave through the gap at $\Phi = 2 \pi$. 
Only after it evolves into an `incidence from the right' state, it will 
eventually escape.

\section{Dependence on the Source Location Inside the Equiangular Spiral}

Depending on where the source {\boldmath $x_s$} of a light ray is situated and
in which direction it is emitted, it will at its first point of reflection 
{\boldmath $x$} either be incident from the right or possibly from the left
relative to $\mbox{\textit{\textbf{\^{x}}}}(\Phi)$. An illustration of this
is given in fig.~\ref{fg:scloc}.   
\bfg
  %\begin{picture}(150,150)
    %\put(65,0){\scalebox{0.4}{\includegraphics[0in,1in][9in,9in]{RLsectors1}}}
		%\put(65,0){\scalebox{0.4}{\includegraphics{RLsectors1}}}
  %\end{picture} 
	\begin{center}
	  \scalebox{0.4}{\includegraphics{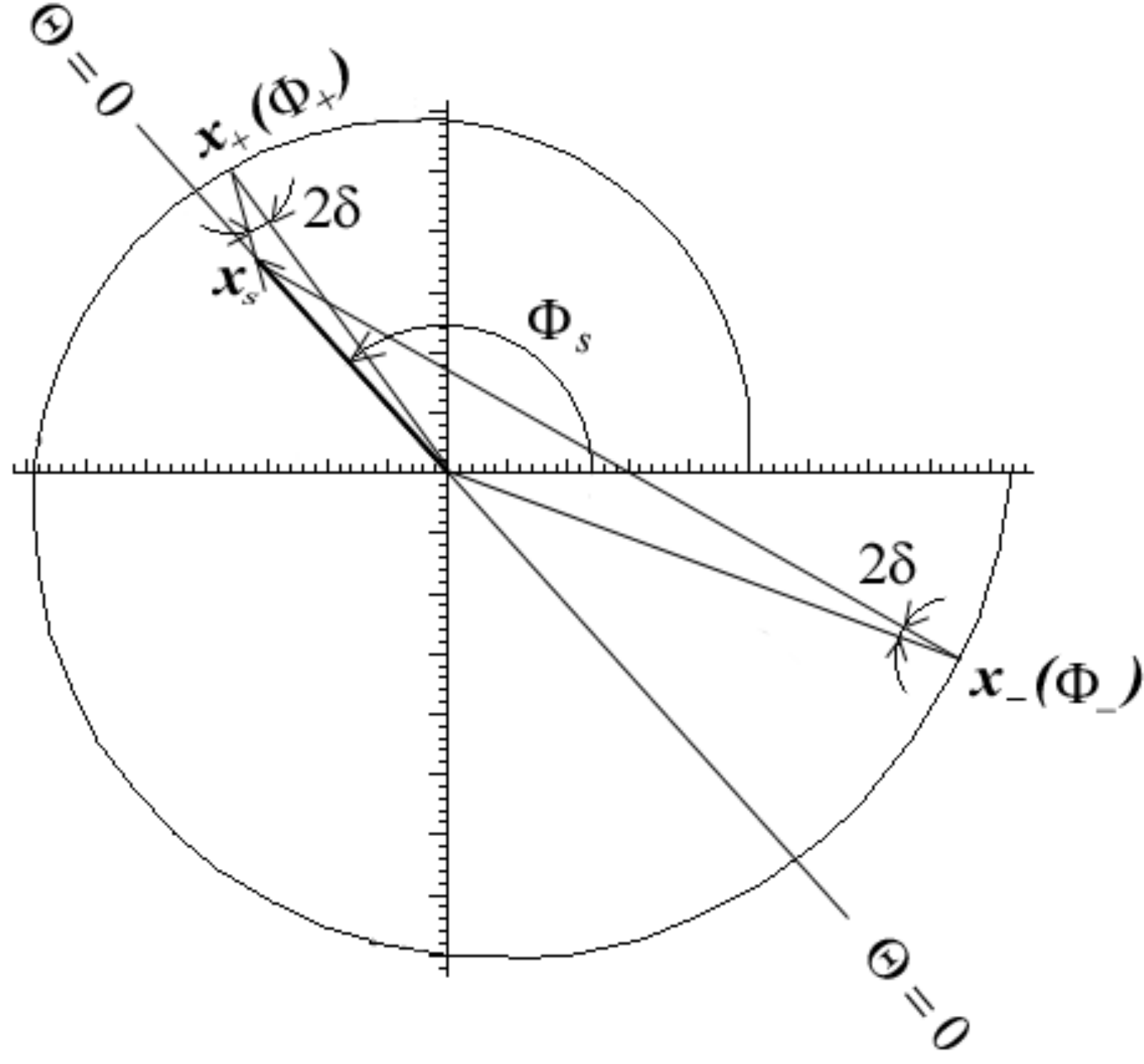}}
	\end{center}
% \vspace{2cm}
  \caption{Sectors of incidence from the right (and left).}
  \label{fg:scloc}
\efg
 Fig.~\ref{fg:scloc} shows, that depending on the source location {\boldmath $x_s$} 
there are in principle two sectors: for all 
$\mbox{\textit{\textbf{{x}}}}(\Phi)$ on the equiangular spiral in the region 
$\Phi \in [\Phi_{+},\Phi_{-}[$ rays emitted from {\boldmath $x_s$} will appear
to be incident from the right relative to $\mbox{\textit{\textbf{\^{x}}}}(\Phi)$,
whilst for all $\mbox{\textit{\textbf{{x}}}}(\Phi)$ on the equiangular spiral with 
$\Phi \in [\Phi_{-},2 \pi[$ or $\Phi \in [0,\Phi_{+}[$ rays emitted from 
{\boldmath $x_s$} will appear to be incident from the left relative to  
$\mbox{\textit{\textbf{\^{x}}}}(\Phi)$. $\Phi_{+}$ and $\Phi_{-}$ will both be functions
of $\Phi_s$ and $x_s$, determined by the condition $\ta = - 2 \delta$.

In the plane geometric algebra $\mathcal{G}_2$ this may be expressed by
\bea
  \mbox{\textit{\textbf{\^{x}}}}_{\pm} {}\mbox{\textit{\textbf{\^{r}}}}_{\pm} 
  & = & exp(-2 \mathbf{i}\delta) 
  \label{eq:source1a}
  \\
  \mbox{where \hspace{1cm} }  \mbox{\textit{\textbf{r}}}_{\pm} 
  & = & \mbox{\textit{\textbf{x}}}_{\pm} - \mbox{\boldmath $x_s$}, 
  \\
  \mbox{\textit{\textbf{{x}}}}_{\pm} & = &
  \mbox{\boldmath $x_0$} exp((\mathbf{i}+t)\Phi_{\pm}) 
  \label{eq:source1c} 
  \\
  \mbox{and \hspace{1cm} } \mbox{\textit{\textbf{{x}}}}_s 
  & = &
  x_s \frac{\mbox{\boldmath $x_0$}}{x_0} exp(\mathbf{i}\Phi_s)
  \nonumber
\eea
Inserting 
$\mbox{\textit{\textbf{\^{r}}}}_{\pm} = \mbox{\textit{\textbf{r}}}_{\pm} / r_{\pm}$
and
$\mbox{\textit{\textbf{\^{x}}}}_{\pm} = \mbox{\textit{\textbf{x}}}_{\pm} / x_{\pm}$
into eq. (\ref{eq:source1a}) we get:
\be
  x_{\pm}^2 - \mbox{\textit{\textbf{x}}}_{\pm} \mbox{\textit{\textbf{x}}}_s
  =  x_{\pm} \mid \mbox{\textit{\textbf{x}}}_{\pm} - \mbox{\textit{\textbf{x}}}_s \mid
     exp(-2 \mathbf{i} \delta)
\ee
divided by $x_{\pm}$ this gives
\be
  x_{\pm} 
  - \frac{\mbox{\textit{\textbf{x}}}_{\pm} \mbox{\textit{\textbf{x}}}_s}{x_{\pm}}
  = \mid \mbox{\textit{\textbf{x}}}_{\pm} - \mbox{\textit{\textbf{x}}}_s \mid 
    exp(-2 \mathbf{i} \delta).
\ee
Inserting now (\ref{eq:source1c}) results in
\bea
 x_0 exp(t \Phi_{\pm}) - \frac{1}{x_o exp(t \Phi_{\pm})} 
                         \mbox{\boldmath $x_0$}
                         exp(\mathbf{i}\Phi_{\pm})
                         exp(t \Phi_{\pm}) \frac{x_s}{x_0} 
                         \mbox{\boldmath $x_0$} 
                         exp(\mathbf{i}\Phi_s) 
&& \nonumber \\
  = \mid \mbox{\textit{\textbf{x}}}_{\pm} - \mbox{\textit{\textbf{x}}}_s \mid 
    exp(-2 \mathbf{i} \delta). && 
\eea
By interchanging {\boldmath $x_0$} and $exp(\mathbf{i} \Phi_{\pm})$ and multiplying both
sides with\linebreak 
$exp(2 \mathbf{i} \delta)/x_0$ we get
\be
  exp(t \Phi_{\pm}) exp(2 \mathbf{i} \delta) - \frac{x_s}{x_0} 
  \frac{\mbox{\boldmath $x_0$}^2}{x_0^2} exp(\mathbf{i}(\Phi_s -\Phi_{\pm}+2 \delta)) 
  = \frac{\mid \mbox{\textit{\textbf{x}}}_{\pm} - \mbox{\textit{\textbf{x}}}_s \mid}{x_0}
\ee
The bivector part of this equation divided by \textbf{i} reads
\be
  exp(t \Phi_{\pm}) sin 2 \delta - \frac{x_s}{x_0} sin(\Phi_s -\Phi_{\pm}+2 \delta) = 0
\ee
which is equivalent to
\be
  exp(t \Phi_{\pm}) 
  = \frac{x_s}{x_0} \frac{sin(\Phi_s -\Phi_{\pm}+2 \delta)}{sin 2 \delta}
  \label{eq:source2a}
\ee
Equation (\ref{eq:source2a}) is a transcendental equation for $\Phi_{+}$ and $\Phi_{-}$ 
depending on
three parameters: $\delta$ the angle between the radial direction 
{\textit{\textbf{\^{x}}}} and the outer normal {\boldmath $n$}, $\Phi_s$ and $x_s$ 
the polar coordinates of the source {\boldmath $x_s$}. A close look reveals, that
eq. (\ref{eq:source2a}) just expresses the law of sinuses in the triangles formed by
the side vectors {\boldmath $x_s$} and {\boldmath $x_{+}$} or {\boldmath $x_{-}$} 
respectively, since the lhs. $exp(t \Phi_{\pm})$ is equal to the relative amplitudes
$\frac{x_{\pm}}{x_0}$.
\bfg
  %\begin{picture}(150,103)
    %\put(-5,130){\scalebox{2}{\includegraphics[0in,1in][9in,9in]{xsphivar1}}}
		%\put(-5,130){\scalebox{2}{\includegraphics{xsphivar1}}}
  %\end{picture} 
	\begin{center}
	  \scalebox{2}{\includegraphics{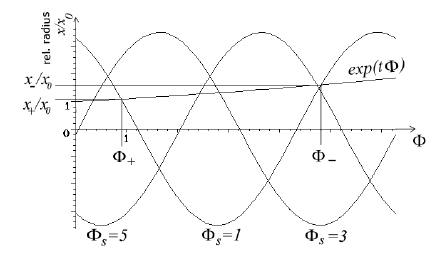}}
	\end{center}
% \vspace{2cm}
  \caption{$\Phi_{+}$ and  $\Phi_{-}$ as functions of the polar angle of the source $\Phi_s$ 
           for constant $\frac{x_s}{x_0}=0.7$ and $\delta = 0.1$.}
  \label{fg:phisdep}
\efg
%%%
%%%
\floatsep2cm
%%%
%%%
Fig.~\ref{fg:phisdep} shows the rhs. of eq. (\ref{eq:source2a}) for constant $x_s$
and varying values of $\Phi_s$. The points of intersection of the sinusoidal curves
with the exponential function $exp(t \Phi)$ gives the desired two pairs of values 
($x_{\pm},\Phi_{\pm}$).
\bfg
  %\begin{picture}(150,103)
    %\put(50,130){\scalebox{2}{\includegraphics[0in,1in][9in,9in]{phixsvar1}}}
		%\put(50,130){\scalebox{2}{\includegraphics{phixsvar1}}}
  %\end{picture} 
	\begin{center}
	  \scalebox{2}{\includegraphics{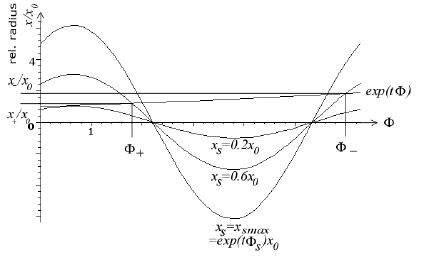}}
	\end{center}
% \vspace{2cm}
  \caption{$\Phi_{+}$ and  $\Phi_{-}$ as functions of the radial location of the source $x_s$
           for constant $\Phi_s= 2 \, rad$ and $\delta = 0.1$.}
  \label{fg:xsdep}
\efg
%%%
%%%
\floatsep1.7cm
%%%
%%%

Fig.~\ref{fg:xsdep} shows the rhs. of eq. (\ref{eq:source2a}) as well, but
this time for fixed $\Phi_s$ and varying $x_s$. 
It can be seen that there will be a certain 
critical value $x_s = x_{cs}$, where the sinussoidal curves representing the rhs. of 
eq. (\ref{eq:source2a}) just touch the lhs. amplitude function $exp(t \Phi)$ only once,
hence $x_{+} = x_{-} = x_{c \pm}$ and $\Phi_{+} = \Phi_{-} = \Phi_{c \pm}$~\footnote{
Please note, that $x_{c \pm}$ and $\Phi_{c \pm}$ each denote only \textit{one unique}
critical value as opposed to the usual $\pm$ notation. }. 
For $x_s < x_{cs}$, all the rays emitted from 
{\boldmath $x_s$} will be incident from the \textit{right} 
at points of the equiangular spiral. 
 
I think therefore that the values of $x_{cs}{(\Phi_s)}$ deserve further attention. 
As it will turn out, an analytical expression for $x_{cs}{(\Phi_s)}$ can be derived. I will
treat this problem first. In the following I will call the set of points given by 
$x_{cs}{(\Phi_s)}$ simply: {\boldmath $x_{cs}$}.

\subsection{The Critical Curve of Source Locations {\boldmath $x_s$}}

The condition for $x_{cs}$ expressed in the above can be mathematically formulated as 
equation (\ref{eq:source2a}) and
\be
  \frac{\partial}{\partial \Phi} exp(t \Phi) \mid_{\Phi = \Phi_{c \pm}}
  = 
  \frac{x_s}{x_0} \frac{1}{sin 2 \delta} 
  \frac{\partial}{\partial \Phi} sin(\Phi_s -\Phi+2 \delta) \mid_{\Phi = \Phi_{c \pm}}
  \label{eq:source2b}
\ee
i.e. the lhs. and the rhs. of eq. (\ref{eq:source2a}) must be equal as well as their
first derivatives with respect to $\Phi$. $\Phi_s$ in eq. (\ref{eq:source2b}) is the  
independent variable upon which $x_s$ depends. 
The value of $x_s$ fulfilling 
eqs. (\ref{eq:source2a}) and (\ref{eq:source2b}) will be the desired $x_{cs}{(\Phi_s)}$.

Performing the differentiation in eq. (\ref{eq:source2b}) and dividing the resulting
lhs. by $exp(t \Phi)$ and the resulting rhs. by the rhs. of (\ref{eq:source2a}) we
obtain
\be
  t = - \frac{1}{tan(\Phi_s + 2 \delta - \Phi_{c \pm})} \mbox{   or   } 
  \frac{-1}{t} = tan(\Phi_s + 2 \delta - \Phi_{c \pm})
\ee 
Using a formula analogous to eq. (\ref{eq:invtan}) we get
\be
  \frac{-1}{t} = tan(\delta -\ph \mp \pi) = tan (\Phi_s + 2 \delta - \Phi_{c \pm})
\ee
Hence
\be
  \Phi_{c \pm} = \Phi_s + \delta + \ph \pm \pi.
\ee
The proper choice of sign turns out to be (comp. fig.~\ref{fg:scrit})
\bea
  \Phi_{c \pm} & = & \Phi_s + \delta +\frac{3}{2} \pi 
  \mbox{   for   } 0 \leq \Phi_s < \ph - \delta
  \nonumber \\
\mbox{and   }
  \Phi_{c \pm} & = & \Phi_s + \delta - \ph 
  \mbox{   for   } \Phi_s \geq \ph - \delta
  \label{eq:phisc}
\eea
Inserting the result (\ref{eq:phisc}) back into eq. (\ref{eq:source2a}) we
obtain 
\be
  x_{cs}(\Phi_s) = 2 x_0 sin \delta \, exp(t(\Phi_s + \delta + \ph \pm \pi))
\ee  
where the plus sign is to be used for $0 \leq \Phi_s < \ph - \delta$ and the
minus sign otherwise.
\bfg  
  %\begin{picture}(150,100)
    %\put(20,350){\scalebox{2.5}{\includegraphics[0in,2in][9in,4in]{criteas1}}}
		%\put(20,350){\scalebox{2.5}{\includegraphics{criteas1}}}
  %\end{picture} 
	\begin{center}
	  \scalebox{2.5}{\includegraphics{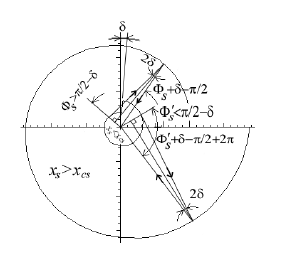}}
	\end{center}
% \vspace{2cm}
  \caption{Critical curve of source locations.}
  \label{fg:scrit}
\efg
Fig~\ref{fg:scrit} shows the critical curve for the source locations $x_s$ 
inside\footnote{
Eq. (\ref{eq:cspiral1}) shows that to say \textit{inside} is actually only justified for sufficiently small values of $\delta$. $\delta$ has
a critical value $\delta_c$ for which part of the critical equiangular spiral will be in congruence 
($0 \leq \Phi < \ph - \delta_c$) with the original equiangular spiral. The condition for $\delta_c$ is that
the factor in the second line of eq. (\ref{eq:cspiral1}) equals one: 
$2 \,sin \,\delta_c \,\,exp(tan\,\delta_c \,\,(\frac{3}{2}\pi+ \delta_c)) \equiv 1\, .$
$\delta_c$ can be numerically determined to be $\delta_c = 0.1929481879\,\, rad,$ 
which is about 11 degrees. For $\delta > \delta_c$ the critical equiangular spiral will 
no longer lay fully inside the original equiangular spiral.} 
the equiangular spiral. That the critical curve itself may again be
an equiangular spiral is already suggested by looking at the figure and confirmed
by writing~\footnote{
It is easy to check that $x_{cs}$ is continuous at $\Phi_s = 0 \,\,\,(2 \pi).$}
\bea
  \mbox{\boldmath $x_{cs}$} & = & \mbox{\boldmath $x_{cs0}$} exp ((\mathbf{i}+t)\Phi_s) 
  \label{eq:cspiral1}
  \\
  \mbox{\boldmath $x_{cs0}$} 
  & = & 2 sin \delta \,\,\, exp(t \Phi_0) \, \mbox{\boldmath $x_0$}
  \nonumber \\
  \Phi_0 & = & \frac{3}{2} \pi + \delta 
  \mbox{   for   } 0 \leq \Phi_s < \ph - \delta 
  \nonumber \\ 
         &   & \mbox{   and   }\,\,\,\, -\ph + \delta    
  \mbox{   otherwise.}
  \nonumber
\eea

This mere fact lends itself to the interesting conclusion, that if one would 
"physically" place an equiangular spiral at the location of the critical equiangular 
spiral (curve) described by eq. (\ref{eq:cspiral1}) inside the original equiangular
spiral given by eq. (\ref{eq:esdef}), all light rays emitted from the gap of this 
critical equiangular spiral would naturally be only right side incident rays 
for points of incidence on the orignal spiral. 

As can be seen from eq. (\ref{eq:cspiral1}) we have 
$\lim_{\delta \rightarrow 0} \mbox{\boldmath $x_{cs}$}= 0$, since 
$\lim_{\delta \rightarrow 0} sin \delta = 0$. This explains the complete absence of any
such critical curve in the case of a circle. 

Another observation is, that for the special case of $x_s = x_{cs}$ 
eq. (\ref{eq:source2a}) can be written as
\be
  \frac{x_{cs}}{sin 2 \delta } = 
  \frac{x_0 exp (t \Phi_{c \pm})}{sin(\Phi_s - \Phi_{c \pm}+ 2 \delta)}
  =
  \frac{x_{c \pm}}{sin(\Phi_s - \Phi_{c \pm}+ 2 \delta)}
  =
  \frac{x_{c \pm}}{sin \Psi}
\ee
with 
\be
  \Psi = \pi - [(\Phi_s - \Phi_{c \pm})+ 2 \delta] 
  = 2(\ph - \delta) - (\Phi_s - \Phi_{c \pm}).
  \label{eq:csangles1}
\ee

From fig.~\ref{fg:scrit} we see that $\Phi - \Phi_{c \pm}$, $\Psi$ and $2 \delta$ are the
three angles of the triangle formed by the origin, {\boldmath $x_{cs}$} and 
{\boldmath $x_{c \pm}$}, where each angle and each point correspond to each other
as listed. In the case of 
$\Phi_s \geq \ph - \delta$ we have $\Phi_{c \pm} = \Phi_s + \delta - \ph$, hence
\be
  \Phi_s - \Phi_{c \pm} = \Phi_s - \Phi_s - \delta + \ph = \ph - \delta 
\ee
We therefore obtain according to (\ref{eq:csangles1}):
\be
  \Psi = \ph - \delta = \Phi_s - \Phi_{c \pm}
  \label{eq:csangles2}  
\ee
This shows that the above mentioned triangle is an equilateral triangle with basis lenght 
$x_{cs}$ and side lenghts $x_{c \pm}$. The same can be shown for 
$0 \leq \Phi_s < \ph - \delta$. 

This in turn can serve as a very simple geometric method in order to construct the 
critical equiangular spiral. For any $\mbox{\boldmath $x$}(\Phi)$ taken as 
$\mbox{\boldmath $x_{c \pm}$}(\Phi)$ one rotates the vector 
$- \mbox{\boldmath $x$}$ attached to {\boldmath $x$} clockwise by $2 \delta$
and obtains the corresponding 
\be
  \mbox{\boldmath $x_{cs}$} = \mbox{\boldmath $x$} + (- \mbox{\boldmath $x$})
  exp(-2 \mathbf{i} \delta) 
  = \mbox{\boldmath $x$} (1- exp(-2 \mathbf{i} \delta) )
  \label{eq:cspiral2}
\ee  

Another interesting property of the critical equiangular spiral follows from 
eq. (\ref{eq:cspiral1}). The $t = tan \delta$ is the same as in the definition 
of the original equiangular spiral (\ref{eq:esdef}). The critical equiangular
spiral is therefore just a shrunk (factor: $2 sin \delta$) and rotated
(anticlockwise by $\ph -\delta$) version of the original one. 

There is in addition even a way to "see" the critical equiangular spiral when
placing a light ray source at the origin. Because according to eq. (\ref{eq:csangles2})
the angle $\Psi$ in fig.~\ref{fg:scrit} is exactly $\ph - \delta$. Comparing
this with eq. (\ref{eq:estangle}) we therefore see that the rays reflected at any
point {\boldmath $x$}(= {\boldmath $x_{c \pm}$}) of the original equiangular spiral
are precisely tangential to the critical equiangular spiral. The set of one 
time reflected rays, which originated at the origin represents therefore
the set of all tangents to the critical equiangular spiral.\footnote{
In other words, the critical equiangular spiral is the \textit{envelope} 
of the set of once reflected rays (regarded as a family of curves), which originated at the origin.}

Knowing this and comparing fig.~\ref{fg:scrit} with fig.~\ref{fg:scloc} shows that 
for $x_s(\Phi_s) >x_{cs}(\Phi_s)  $ the two straight lines 
through {\boldmath $x_{s}$} and {\boldmath $x_{+}$}, and through {\boldmath $x_{s}$} 
and {\boldmath $x_{-}$} in fig.~\ref{fg:scloc} are actually the
two tangents of the critical equiangular spiral through {\boldmath $x_{s}$}. This gives a simple geometric construction
in order to find the two angular sectors of left- and right incidence for any source {\boldmath $x_{s}$} with the help of 
the critical equiangular spiral.

\section{Conclusion}

In this work, I first introduced the way in which geometric calculus describes
an equiangular spiral and reflections of light at it. I then discussed the development
of a light path through successive reflections. I made a distinction between 
incidence from the left and from the right relative to the radius vector. 
It was then found that right incident vectors continue to be right incident,
and follow anticlockwise polygonal paths bending closer and closer to the 
boundary until they eventually leave through the gap. 

As for left incident light rays, they first follow clockwise polygonal paths, 
yet bending further and further away from the boundary until they eventually 
change their \textit{state} into right incident rays with anticlockwise 
polygonal paths. 

The conditions of escape and the asymptotic characteristics were analyzed in detail
as well. 

Finally point sources were placed inside the spiral and the occurence of right- and
left incidence, depending on the source location was examined. The very interesting
structure of another equiangular sprial, a dilated and rotated concentric 
version of the original one, was discovered. Rays emitted from any source inside 
this \textit{critical} equiangular spiral area were found to be right incident 
rays at the original equiangular spiral ab initio. 
The discussion of this critical equiangular spiral was concluded with
explaining some of its geometrical and physical properties. 

This short treatment of the subject may in itself only be an introduction. Further
analysis may still be very fruitful. One may e.g. ask for higher dimensional 
spirally deformed objects or cavities, like as cones and spheres. 
Or one may try to impose spiral deformations not only on circles but on other
conical sections as well. One may further ask, whether the phenomenon of the
self similar second critical equiangular spiral is an artifact of the globally
constant $\delta$, or if corresponding structures would exist for $\delta$ being
a function of $\Phi$ as well.

The application of numerical techniques may yield a variety of results as well, since
in this work I so far have not yet introduced such methods. The transcendental 
character of the equations will certainly make it necessary for obtaining
more quantitative results. 

The potential applications of this sort of analysis should be very wide, including
the fields of optics, electromagnetic waves in general as well as acoustics. 

In optics, it may lead amongst other possible applications, 
to the development of new types of laser resonators\footnote{Results on potential laser modes
in equiangular spiral cavities will be published in a later paper.} and to 
logical components for optical computing~\footnote{
Left incident and right incident states may both be selected at will by 
introducing suitably wound spiral structures.}. 
New types of telescopes may arise as well from further analyzing image propagation
(i.e. ensembles of light rays defining images). 

For electromagnetic waves in general, new types of resonators and antennas may result.

As for acoustics, architects might find it interesting to construct spirally shaped
buildings, instead of circular ones in order to conduct the sound in certain directions.

\section{Acknowledgements}

I first of all want to thank God, my creator, who allowed me to enjoy this 
interesting piece of research~\cite{GOD:psalm111} in the first place. An ancient biblical acrostic
poem declares: \textit{"Great are the works of the} \textsc{Lord}\textit{; They are pondered by all who delight in them. ... 
The fear of the} \textsc{Lord} \textit{is the beginning of wisdom; all who follow his precepts have good understanding. 
To him belongs eternal praise."} I am indebted to K. Shinoda, who always encouraged me and prayed
for me during this research.
I further thank J.S.R. 
Chisholm, who greatly stirred my interest in geometric calculus. Discussions
with H. Ishi at the university of Kyoto proved very helpful for advancing the ideas
presented. I finally thank the university of Fukui for providing the environment for
carrying out this research work.

%\begin{picture}(10,10)
% \put(0,0){\rotatebox{270}{\includegraphics[3.4in,4.9in][5.1in,6.1in]{aHATb.bmp}}}
%\end{picture}

\bibliography{garef}

\end{document}